\documentclass[twocolumn,fleqn,usenatbib,letters]{mnras}

\usepackage{amssymb}
\usepackage{amsmath}

\usepackage{graphicx}
\usepackage{graphics}
\usepackage{color}
\usepackage{fancyhdr} 
\usepackage{graphicx}
\usepackage{float}
\usepackage[dvipsnames]{xcolor}
\usepackage{comment}

\usepackage{subfig}

\usepackage[utf8]{inputenc}

\usepackage[justification=centerlast, format=plain, labelfont=bf]{caption}

    \newcommand{\be}{\begin{equation}}
  \newcommand{\ee}{\end{equation}}
    \newcommand{\ba}{\begin{align}}
  \newcommand{\ea}{\end{align}}

\newcommand{\Msun}{M_{\odot}}

\newcommand{\MUV}{M_{\rm UV}}
\newcommand{\sUV}{\sigma_{\rm UV}}

\title{Breaking degeneracies in the first galaxies with clustering}

\author[J.B.~Mu\~noz et al.]{
Julian B.~Mu\~noz,$^{1}$\thanks{E-mail: julianbmunoz@utexas.edu}
Jordan Mirocha,$^{2,3}$
Steven Furlanetto,$^{4}$ and
Nashwan Sabti$^{5}$
\\
$^{1}$Department of Astronomy, The University of Texas at Austin, 2515 Speedway, Stop C1400, Austin, TX 78712, USA \\
$^{2}$Jet Propulsion Laboratory, California Institute of Technology, 4800 Oak Grove Drive, Pasadena, CA 91109, USA \\
$^{3}$California Institute of Technology, 1200 E. California Boulevard, Pasadena, CA 91125, USA \\
$^{4}$Department of Physics \& Astronomy, University of California, Los Angeles, Los Angeles, CA 90095, USA \\
$^{5}$William H. Miller III Department of Physics and Astronomy, 3400 N. Charles St., Baltimore, MD 21218, USA \\
}

\date{Accepted XXX. Received YYY; in original form ZZZ}

\pubyear{2023}

\begin{document}
\label{firstpage}
\pagerange{\pageref{firstpage}--\pageref{lastpage}}
\maketitle

\begin{abstract}
The high-redshift galaxy UV luminosity function (UVLF) has become essential for understanding the formation and evolution of the first galaxies.
Yet, UVLFs only measure galaxy abundances, giving rise to a degeneracy between the mean galaxy luminosity and its stochasticity.
Here, we show that upcoming clustering measurements with the James Webb Space Telescope (JWST), as well as with Roman, will be able to break this degeneracy, even at redshifts $z \gtrsim 10$.
First, we demonstrate that current Subaru Hyper Suprime-Cam (HSC) measurements of the galaxy bias at $z\sim 4-6$ point to a relatively tight halo-galaxy connection, with low stochasticity.
Then, we show that the larger UVLFs observed by JWST at $z\gtrsim 10$ can be explained with either a boosted average UV emission or an enhanced stochasticity.
These two models, however, predict different galaxy biases, which are potentially distinguishable in JWST and Roman surveys.
Galaxy-clustering measurements, therefore, will provide crucial insights into the connection between the first galaxies and their dark-matter halos, and identify the root cause of the enhanced abundance of $z \gtrsim 10$ galaxies revealed with JWST during its first year of operations.
\end{abstract}


\maketitle

\section{Introduction}

The ultraviolet luminosity function (UVLF) has emerged as a crucial observational tool to understand the formation of the first galaxies during cosmic dawn and reionization~\citep[e.g.,][]{Finkelstein_2015_review}.
Analyses of the UVLF have derived important constraints on the evolution of high-redshift galaxies~\citep{Trenti:2010sz,Ceverino_FirstLight_UVLF,Tacchella:2018qny} and the process of cosmic reionization~\citep{Mirocha_UVLFs_2017,Park:2018ljd,Mason:2019oeg}, as well as set new cosmological bounds~\citep{Corasaniti:2016epp,Menci:2018lis,Sabti:2020ser,Rudakovskyi:2021jyf,Sabti:2021unj, Sabti:2023xwo}.
In practice, fitting the UVLFs has become a litmus test for any model of high-$z$ galaxy formation.

Deep imaging surveys with space-based telescopes, such as the {\it Hubble} and {\it James Webb} Space Telescopes (HST and JWST, respectively), have extended the reach of UVLFs to ever-increasing redshifts, pushing the observational frontiers closer to the cosmic dawn~\citep{Livermore:2016mbs,Atek:2015axa,Bouwens:2014fua,Finkelstein_CEERS,Treu:2022iti,Eisenstein_JADES}.
Intriguingly, the abundance of star-forming galaxies at $z\gtrsim 10$, reported from the first JWST observations in~\citet{Finkelstein_2022_Maisies,Harikane_UVLFs,Donnan_UVLFs_2023,Bouwens:2022gqg,Finkelstein_CEERS,Castellano_GLASS_hiz,Naidu2022,Leung_NGdeep_2023,Adams_Conselice_JWST_2023}, appears at odds with expectations: there seems to be significantly more UV-bright galaxies than predicted by most galaxy-formation models.
While only a subset of candidates have been spectroscopically confirmed so far~\citep{Fujimoto:2023orx,Arrabal_CEERS_spectra,Curtis-Lake2022:JWST,Harikane_specz_UVLF_JWST,Kocevski_CEERS_spectra_AGN}, with the highest-redshift one identified to be a contaminant~\citep{Arrabal_CEERS_spectra,Zavala_2023_interlopers}, there has been a flurry of theoretical activity to explain the photometrically derived UVLFs, with solutions broadly coming in two flavors: either an enhanced star-formation/UV emission at high $z$~\citep{Ferrara2022,Inayoshi_JWST_2023,Dekel:2023ddd,Yung:2023bng,Steinhardt_PopIII}, or a large ``stochasticity'' in the UV brightness of the first galaxies~\citep{Mason:2022tiy,Mirocha_UVLFs2023,Shen:2023cva,Padmanabhan:2023esp}.
Both models are able to explain the JWST data by producing more UV-bright galaxies, in the former case as an average and in the latter as the tail of a distribution.

\begin{figure}
    \centering
\includegraphics[width=\linewidth]{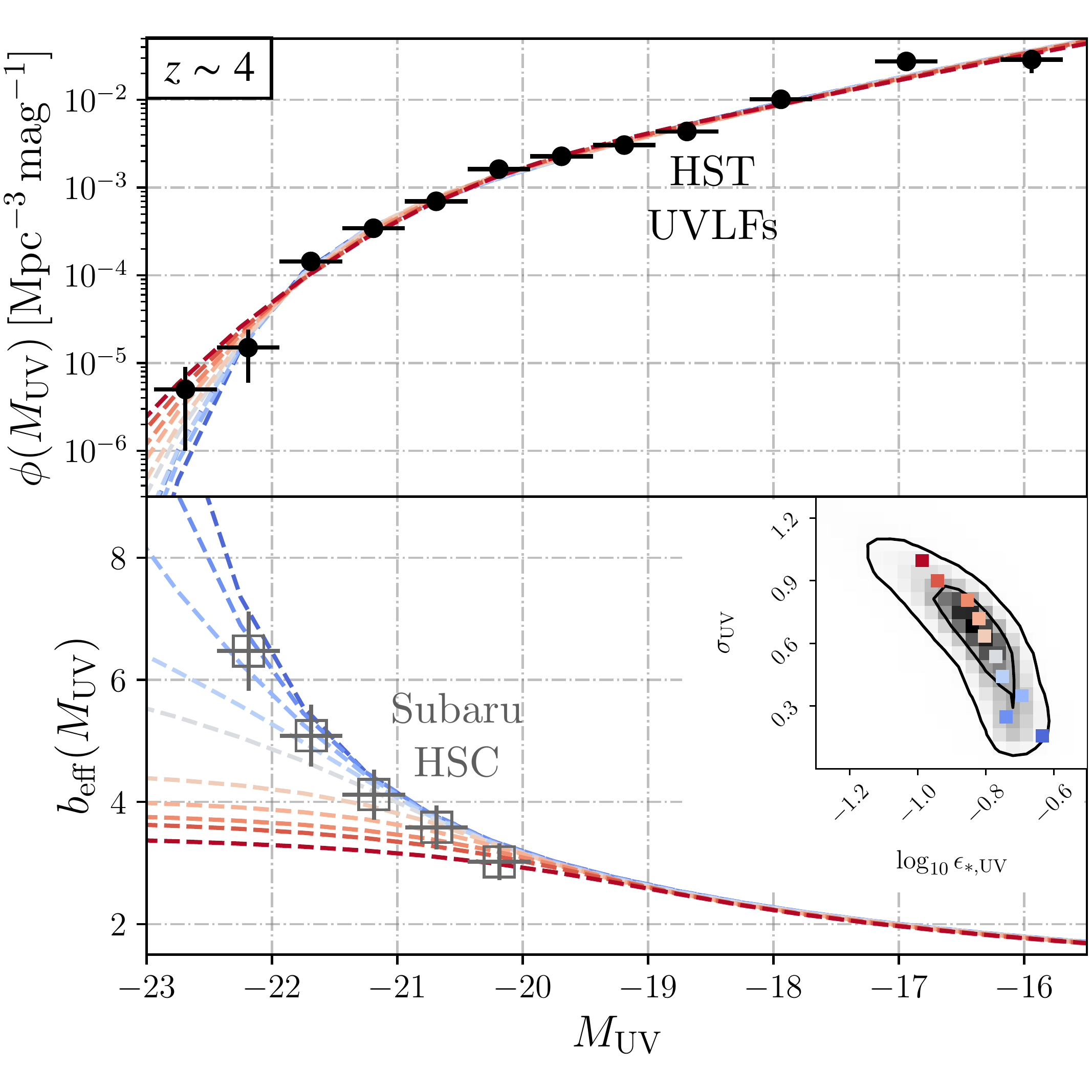}
    \caption{Illustration of the degeneracy between stochasticity and star-formation efficiency at high $z$. We show the UVLF ({\bf top}) and the bias ({\bf bottom}) at $z\sim 4$ as a function of UV magnitude for different models.
    These are all calibrated to the UVLF data (black circles, from~\citealt{Bouwens_2021_UVLFs}) and colored by their standard deviation $\sigma_\mathrm{UV}$ (in the $P(M_\mathrm{UV}|M_\mathrm{h})$ distribution), whose posterior against the star-formation efficiency $\epsilon_{\star,\rm UV}$ is shown in the inset.
    The bias predictions differ dramatically, depending on whether heavy or light halo masses are responsible for the bright end of the UVLF. 
    We show the biases reported in~\citet{Harikane_Goldrush_2021} as gray open squares, which have not been used in the inference.
    All predictions are binned with $\Delta M_{\rm UV}=0.5$.
    }
    \label{fig:z4degeneracy}
\end{figure}

This exemplifies an {\it intrinsic degeneracy} in the UVLFs.
UVLFs measure the abundance of galaxies, which would would be sufficient for high-$z$ studies if galaxy luminosities were linked one-to-one to their host dark-matter halo properties~\citep{Behroozi:2010rx,Moster:2012fv}.
However, a more complex---or stochastic---halo-galaxy connection gives rise to more than one way to obtain the same average density of galaxies~\citep{Ren_2019_scatter,Mirocha:2020bias}.
We illustrate this degeneracy in Fig.~\ref{fig:z4degeneracy}, where we show an array of models that can explain the observed UVLF at $z\sim 4$ (from \citealt{Bouwens_2021_UVLFs}), but have vastly different halo-galaxy connections (with stronger or weaker stochasticity).

Here we argue that galaxy clustering can break this degeneracy at high redshifts, akin to halo-occupation-distribution (HOD) studies at lower $z$~\citep[e.g.,][]{Giavalisco:2000xs,Kravtsov:2003sg,Zheng:2005ef,Moster:2009fk,Zentner_assemblybias,Hadzhiyska:2021kmt}, and that this clustering can be measured by current-generation surveys.
Galaxies residing in heavier dark-matter halos will cluster more strongly than their lighter counterparts, boasting a larger galaxy bias.
As such, measuring clustering at high $z$ will allow us to determine whether UV-bright galaxies 
are rare because their host halos are rare as well, or because they only form stars efficiently some fraction of the time.
To illustrate this point, we will show how current bias data at $z\sim4-6$ tend to prefer a closer halo-galaxy connection (with lower stochasticity, and a duty cycle near unity).
Turning to $z\gtrsim 10$, we will argue that upcoming surveys, like Cosmos-Web with JWST~\citep{Casey:2022amu} and the {\it Roman} high-latitude survey (HLS,~\citealt{Spergel:2015sza}), will be able to distinguish between models of galaxy formation.
In particular, they will help determine whether the first galaxies showed enhanced stochasticity or higher UV emission.
Throughout this work, we will fix our cosmological parameters to $h=0.7$ and $\Omega_\mathrm{m}=0.3$ to match~\citet{Bouwens_2021_UVLFs}, set $\sigma_8 = 0.85$, $n_\mathrm{s} = 0.966$, $\omega_\mathrm{b} = 0.022$, and use AB magnitudes~\citep{Oke_Gunn_AB_mags}.

\section{The degeneracy in the UVLFs}
\label{sec:degeneracy}

Let us begin by laying out our definitions, and with them the intrinsic degeneracy in the UVLF.
The UVLF measures the number density of objects with a particular UV magnitude $\MUV$. 
Assuming all galaxies live in dark-matter halos (and a halo-occupation fraction of unity), 
we can find the UVLF as:
\be
\phi_{\rm UV} \equiv \dfrac{\mathrm{d}n}{\mathrm{d} \MUV} = \int \mathrm{d}M_\mathrm{h} \dfrac{\mathrm{d}n}{\mathrm{d}M_\mathrm{h}} P(\MUV | M_\mathrm{h})\ ,
\ee
where $\mathrm{d}n/\mathrm{d}M_\mathrm{h}$ is the halo mass function, following the~\citet{Sheth:2001dp} fit (see also~\citealt{Rodriguez-Puebla:2016ofw}), and $ P(\MUV | M_\mathrm{h})$ is the probability that a halo of mass $M_\mathrm{h}$ hosts a galaxy with magnitude $\MUV$.
The UVLF is often reported in bins:
\be
\phi_{\rm UV}^{\rm bin} = \int \mathrm{d}z\, W_z(z) \int \mathrm{d}\MUV W_{\rm UV}(\MUV) \phi_{\rm UV}(\MUV,z)\ ,
\label{eq:binning}
\ee
where both window functions $W_i$ integrate to unity.
The magnitude windows $W_{\rm UV}$ are assumed to be tophats, whereas the $W_z$ are Gaussians fit to the redshift distributions in~\citet[][where it is important for data and predictions to match the true mean and width $z$, see also~\citealt{Trapp:2021ufe}]{Bouwens_2021_UVLFs}.

The $P(\MUV | M_\mathrm{h})$ term, despite its apparent simplicity, encodes the complex halo-galaxy connection~\citep{Wechsler:2018pic}, including any stochasticity.
We will model it through a semi-analytic approach where $P(\MUV | M_\mathrm{h})$ is a Gaussian centered around a predicted ``mean" $\overline{\MUV}(M_\mathrm{h})$, with a mass-independent dispersion $\sigma_{\rm UV}$ that we allow to vary.
The UV magnitude of a galaxy will depend on its star-formation rate (SFR):
\be
\dot M_\star= f_\star(z, M_\mathrm{h}) f_\mathrm{b} \dot M_\mathrm{h}\ ,
\ee
where $f_\mathrm{b}\approx 0.16$ is the baryon fraction~\citep{Planck:2018vyg}, and we take a model of exponential accretion $M_\mathrm{h}(z) \propto e^{a_{\rm acc} z}$ with $a_{\rm acc}=0.79$~\citep[][see also App.~\ref{app:accretionrates}]{Schneider:2020xmf}.
The average halo-galaxy connection is encoded in the star-formation efficiency (SFE) $f_\star$, generically a function of redshift $z$ and halo mass $M_\mathrm{h}$.
Inspired by analytic and simulation studies~\citep{Moster:2009fk,Furlanetto2016:feedback}, we assume a double-power law functional form:
\be
f_\star = \dfrac{2 \epsilon_\star}{ (M_\mathrm{h}/M_c)^{-\alpha_\star} + (M_\mathrm{h}/M_c)^{-\beta_\star}}\ ,
\label{eq:fstar}
\ee 
which was shown to fit well both observations and simulations in~\citet{Sabti:2021xvh}.
This $f_\star$ has four free parameters: an amplitude $\epsilon_\star$, a critical mass $M_c$, and two power-law indices $\alpha_\star>0$ and $\beta_\star<0$ for the faint  and bright ends, respectively.
The SFR is converted to UV luminosity through $L_{\rm UV} = \dot M_\star/\kappa_{\rm UV}$. Given that the conversion factor $\kappa_{\rm UV}$ is fully degenerate with $\epsilon_\star$, we will define:
\be
\epsilon_{\star,\rm UV} \equiv \epsilon_\star \left(\kappa_{\rm UV}/\overline{\kappa_{\rm UV} }\right)^{-1}\ ,
\ee
for a fiducial $\overline{\kappa_{\rm UV}} = 1.15 \times 10^{-28}$ ($\Msun\,\rm yr^{-1}$) /(erg s$^{-1}$) as in~\citet{Madau:2014bja}.
The UV luminosity per unit SFR could be higher for a top-heavy initial mass function (as expected for Population III stars,~\citealt{Bromm2001_firststars}), and varying $\epsilon_{\star,\rm UV}$ will allow us to capture such behavior.
In addition, we ought to account for dust attenuation, which reduces the apparent brightness of galaxies, especially towards lower $z$ and the bright end~\citep{Yung_2018}.
Following~\citet{Sabti:2021xvh}, we adopt the IRX-$\beta$ dust prescription calibrated by~\citet{Meurer:1999jj} with the $\beta$ measured in~\citet{Bouwens:2013hxa}.
We extrapolate to $z>8$ by using the $z=8$ result, as advocated in~\citet{Mason:2022tiy}.

While simple, this semi-analytical formalism is highly predictive, though it is not without caveats.
One has to allow the free parameters to vary with $z$ to capture the time-dependent impact of feedback, as well as float $\sUV$ against $\MUV$ to study its luminosity dependence.
In this first work we will account for the former effect only.
We fit the five free parameters (four in the $f_\star$ relation plus $\sigma_{\rm UV}$) to the $z=4$ UVLF data from~\citet{Bouwens_2021_UVLFs}\footnote{To account for cosmic variance, we impose a minimum 20\% error in the UVLF data~\citep{Sabti:2021xvh}.}, and show in Fig.~\ref{fig:z4degeneracy} the results for an array of parameters within the 2$\sigma$ preferred region, colored by their stochasticity $\sUV$.
From this figure it is clear that there is a degeneracy between the {\it average} halo-galaxy connection $f_\star$ and its stochasticity $\sUV$, most obvious in the $\epsilon_{\star,\rm UV}-\sUV$ plane.
In essence, there are two ways to explain the paucity of objects in the bright end of the UVLF.
One option is that those galaxies reside in very massive (and therefore rare) halos in the early universe.
The other is that these galaxies are hosted in lower-mass halos, but correspond to the high-brightness tail of a stochastic halo-galaxy connection.
This degeneracy is intrinsic to one-point functions, like the UVLFs.

Here we argue that these two scenarios (which can be rephrased in terms of a duty cycle $f_{\rm duty}$, see~\citealt{Lee:2005jha} and our App.~\ref{app:stochfduty}), are distinguishable by using clustering information, such as two-point functions, as massive halos cluster more strongly than lighter ones.
In this work, we will examine clustering via the halo bias $b(M_\mathrm{h})$, the linear ratio of the galaxy and matter densities~\citep[][though of course more information can be gleaned through nonlinear studies;~\citealt{Berlind:2001xk,Cooray:2002dia}]{Desjacques:2016bnm}, fitted in~\cite{Tinker:2010my}.
Using our formalism, we calculate the number-weighted effective bias as:
\be
b_{\rm eff}(\MUV) = \phi_{\rm UV}^{-1}  \int \mathrm{d}M_\mathrm{h} \dfrac{\mathrm{d}n}{\mathrm{d}M_\mathrm{h}} b(M_\mathrm{h}) P(\MUV | M_\mathrm{h})\ ,
\ee
which we bin as in Eq.~\eqref{eq:binning} and show for each of our models in the bottom panel of Fig.~\ref{fig:z4degeneracy}.
Models with large stochasticity ($\sUV$) 
predict a very mild dependence of bias against intrinsic magnitude, as the bright end will be dominated by smaller-mass objects that have been upscattered in luminosity.
Conversely, a tighter halo-galaxy connection (lower $\sUV$) predicts a sharp rise in the bias towards the bright end, as those objects reside in more massive (and thus exponentially suppressed) halos.
The behavior of $b_{\rm eff}$ is therefore critical to break degeneracies in the halo-galaxy connection, especially towards the bright end.

Along with our predictions, Fig.~\ref{fig:z4degeneracy} shows bias measurements as reported in~\citet{Harikane_Goldrush_2021}, which were obtained from clustering studies with the {\it Subaru} Hyper Suprime-Cam (HSC) at $z=4$.
While adding these measurements to our likelihood is tempting, a word of caution is warranted.
The HSC biases were inferred for several magnitude cuts using an HOD model fitted within a different fiducial cosmology than ours.
As such, we converted these measurements for illustration purposes only (see App.~\ref{app:biasHSC} for our procedure), 
and show them with a 10\% noise floor to account for these uncertainties, leaving a direct likelihood analysis to future work.
Nevertheless, it is clear that the bias measured with HSC rises sharply towards brighter magnitudes, which indicates a preference for  lower stochasticity ($\sUV\lesssim 0.5$ in our model).
In~\citet{Harikane_Goldrush_2021}, this appeared as a close connection in the $M_\mathrm{h}-M_{\rm UV}$ plane (e.g., their Fig.~16).
We refer the interested reader to App.~\ref{app:z5to7}, where we repeat this analysis by simultaneously varying $\sUV$ and $\epsilon_{\star,\rm UV}$ at each redshift slice in the $z=5-7$ range, finding similar results.

It is then clear that galaxy clustering has the potential to break the degeneracy between the average halo-galaxy connection ($f_\star$) and its stochasticity ($\sUV$).
Let us now study the impact of galaxy clustering in  light of the $z\gtrsim 10$ JWST data.

\section{Fitting the high-redshift UVLFs}
\label{sec:hizgals}

So far, we have shown that at a single redshift there is an intrinsic degeneracy in the UVLFs between the amplitude of the SFE of galaxies and their stochasticity.
Of course, if these two parameters had a known $z$ evolution, co-adding information from different redshifts could help break this degeneracy.
There is, however, no guarantee that either $\epsilon_{\star,\rm UV}$ or $\sUV$ are constant with $z$.
For instance, the shape of the SFE $f_\star$ can evolve with the strength of astrophysical feedback~\citep{Furlanetto2016:feedback}.
Even with a fixed SFE, galaxies become more metal-rich over time, which affects the $\kappa_{\rm UV}$ conversion between SFR and UV emission (and thus $\epsilon_{\star,\rm UV}$). 
Likewise, $\sUV$ could evolve as reionization and feedback strip gas away from small-mass galaxies~\citep{Faucher_2018_feedback,Furlanetto_Mirocha_21_bursty}.
Rather than fixing these parameters, we will use two simple models to determine their behavior as a function of redshift.
In the first, we will vary $\sUV$ independently at each redshift, and vary the rest of parameters linearly with $z$ (specifically, the two power-law indices $\alpha_\star$ and $\beta_\star$, as well as $\log_{10}M_c$ and  $\log_{10} \epsilon_{\star, \rm UV}$).
In the second model, we will switch the roles of $\log_{10} \epsilon_{\star, \rm UV}$ and $\sUV$, making the latter linear in $z$ and the former vary at each slice.

\begin{figure*}
    \centering
\includegraphics[width=0.9\linewidth]{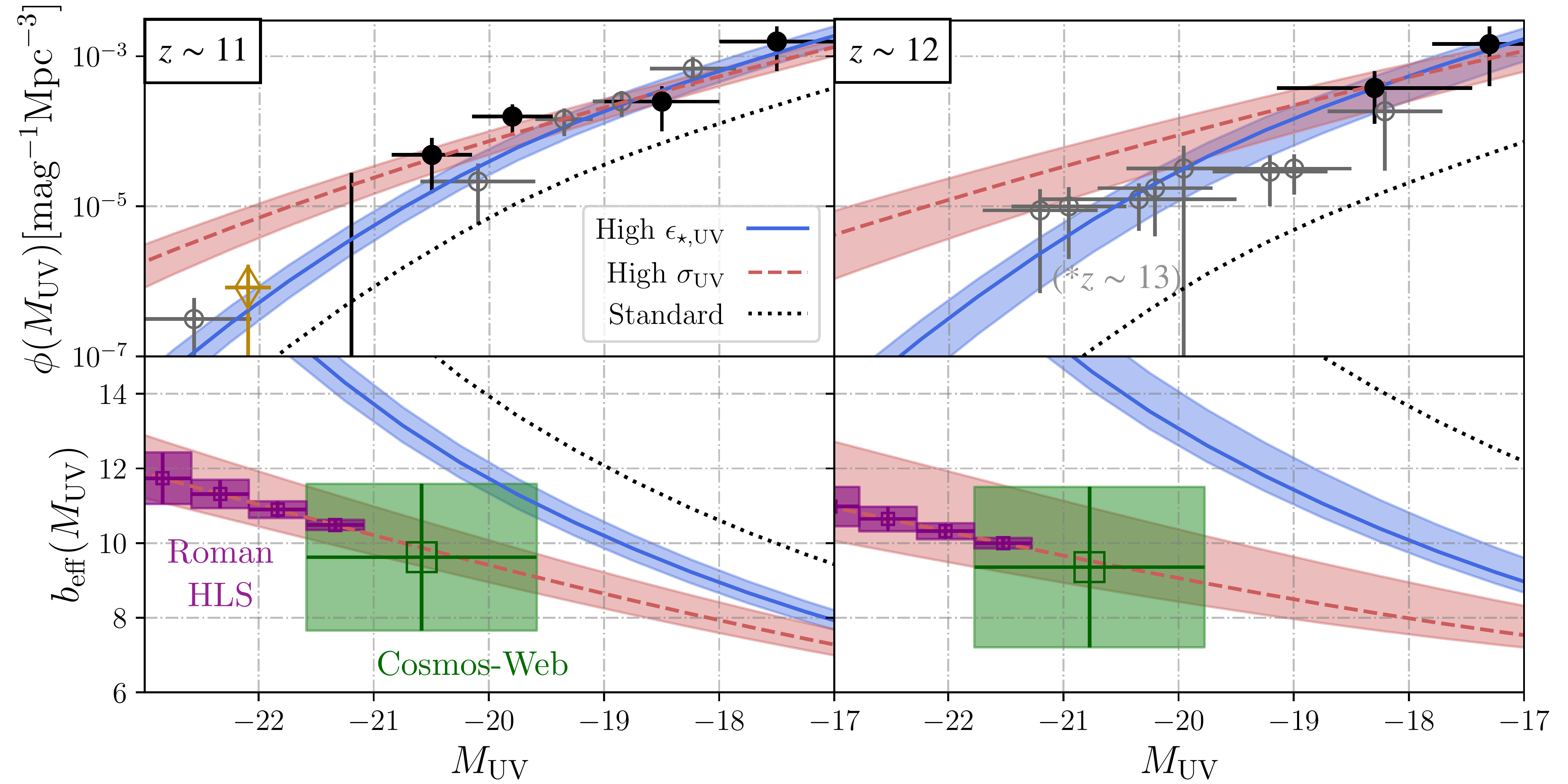}
    \caption{UVLFs and biases at $z\sim11$ ($z=10.75$ with a $0.5$ rms) and $12$ ($z=12.25$ and $0.75$ rms). Filled black points correspond to the UVLF data used in our analysis, from~\citet{Finkelstein_CEERS,Perez-Gonzalez:2023wta,Harikane_UVLFs}, whereas gray points are data from~\citet{Donnan_UVLFs_2023,Bouwens:2022gqg}, which correspond to $z\sim10.5$ and $z\sim 13$, as indicated.
    The bright ($\MUV < -21$) end of the UVLF at $z\sim 11$ contains two objects: GN-z11~\citep[as a yellow diamond,][]{Gnz11_Oesch}, recently argued to host an AGN~\citep{Maiolino:2023zdu}, and an object that could be an interloper at $z\sim 3$~\citep{Cosmos2020_Kauffman}, which we do not use in our inference.
    We show the 1$\sigma$ predictions for two models, in red for a enhanced stochasticity $\sigma_{\rm UV}(z)$ and in blue for a larger mean UV emission $\epsilon_{\star, \rm UV}(z)$. The black-dotted line is a simple extrapolation of the $z\leq10$ fit from HST, which does not reach the values necessary to explain the JWST UVLF.
    Bottom panels show the bias expected in each model, as well as projected measurements for a Cosmos-Web-like survey and the {\it Roman} high-latitude survey (HLS).
    }
    \label{fig:highz_uvlf}
\end{figure*}

We begin by fitting the HST UVLFs compiled in~\citet{Bouwens_2021_UVLFs}.
These are derived from blank-field surveys in the Legacy Fields catalog covering $z=4-10$ and do not include lensing fields.
We will fit both models, as our benchmarks, with a (log) likelihood at each redshift:
\be
-\log\mathcal L(z) = \sum_{i}  \dfrac{(\phi_{{\rm UV},i}^{\rm obs}-\phi_{{\rm UV},i}^{\rm bin})^2}{2 (\sigma^{\rm obs}_{i})^2}\ ,
\ee
given UVLF measurements $\phi_{{\rm UV},i}^{\rm obs}$ with uncorrelated Gaussian errors $\sigma_i^{\rm obs}$, and our predicted $\phi_{{\rm UV},i}^{\rm bin}$, where the sum runs over magnitude bins $i$.
We then sum the log-likelihoods for all the $z$ we consider (implicitly assuming they are uncorrelated as well).
We run a Markov Chain Monte Carlo for both models and find good fits in each case, with $\chi^2=33.9$ and 34.6 for the variable-$\sUV$ and $\epsilon_{\star, \rm UV}$ models, respectively. This is with 46 degrees of freedom, indicating perhaps an overestimation of the error bars due to the 20\% minimum error imposed to account for cosmic variance.
The $\Delta \chi^2 = 0.7$ difference between the two scenarios shows that both are equally good fits to the HST data.

These two runs, calibrated at $z\leq 10$, are designed to provide a ``standard" against which to compare the JWST data. As we will see, a naive extrapolation to $z\geq 10$ greatly underpredicts the amount of galaxies observed in JWST.
Let us quantify this statement.

The JWST data are not yet constraining enough on their own, so we will leverage the $z\leq 10$ information from HST.
In order to minimize covariances and avoid double-counting, we will only include JWST galaxy candidates above $z=10$.
In particular, we build the UVLF at $z\sim11$ and 12 using the results from CEERS~(\citealt{Finkelstein_CEERS} and~\citealt{Perez-Gonzalez:2023wta}, which can be co-added as the latter only covers fainter objects than the former), and at $z\sim 13$ using data from~\citet{Harikane_UVLFs}.
These are shown in Fig.~\ref{fig:highz_uvlf}, along with the UVLFs from~\citet{Bouwens:2022gqg,Donnan_UVLFs_2023}. We note that we do not use any $z\sim 16$ data as they are currently highly uncertain, 
and take the appropriate average of asymmetric error bars.
Our UVLF data does not contain the two brightest objects reported at $z\sim 11$ (shown as empty symbols in Fig.~\ref{fig:highz_uvlf}).
They correspond to GN-z11~\citep{Gnz11_Oesch}, which has recently been argued to host an active galactic nucleus~\citep[AGN, ][]{Maiolino:2023zdu}, and a galaxy previously reported in COSMOS2020 (ID 1356755) that has secondary solutions at $z\sim 3$~\citep{Cosmos2020_Kauffman}.
These uncertainties underscore the necessity for beyond-UVLF measurements to break degeneracies, as the bright tail of the UVLF is susceptible to contamination from interlopers~\citep{Furlanetto_purity_JWST} and AGN~\citep{Finkelstein_Bagley_22}.

In practice,  we first fit both our models to the HST ($z=4-10$) data, as outlined above, and fix all parameters to the best fit of that analysis.
This then allows us to easily vary the parameter of interest (either $\sUV$ or $\epsilon_{\star,\rm UV}$) at higher $z$ without altering the good fit to low-$z$ HST data.
Fig.~\ref{fig:highz_uvlf} shows the UVLF data at $z\sim 11$ and 12, along with the predictions of the two models (varying $\sUV$ and $\epsilon_{\star,\rm UV}$ at these $z$ within their 1$\sigma$ region), both of which provide a good fit to the JWST data.
In particular, the $\chi^2$ difference between the models (co-adding $z\sim 11 - 13$ data) is $\Delta \chi^2=2.5$, weakly favoring the $\sUV$ model.
Both have $\Delta \chi^2\approx 15$ compared to the standard prediction, obtained by extrapolating from $z\leq10$ (also shown in Fig.~\ref{fig:highz_uvlf}).
Thus, the $z>10$ UVLFs appear to prefer a new component in our models, either large stochasticity or UV brightness.

\begin{figure}
    \centering
\includegraphics[width=\linewidth]{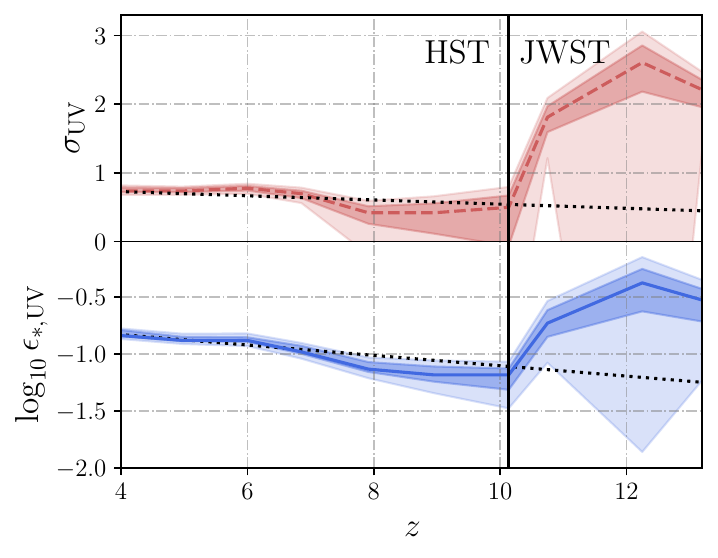}
    \caption{Posteriors for the two models we consider: one with a $z$-dependent stochasticity $\sigma_{\rm UV}$ ({\bf top}), and one with an enhanced UV emission $\epsilon_{\star,\rm UV}$ ({\bf bottom}).
    In each panel, we fix the rest of parameters to their best-fit HST values, and show 1$\sigma$ and 2$\sigma$ contours in dark and light shades, respectively.
    Dotted lines correspond to the best-fit linear evolution in $z$ for each model, which is dominated by the (more constraining) low-$z$ data.
    While the $z \leq 10$ data has a smooth behavior in both variables, the $z=11-13$ data from JWST prefers a jump towards high stochasticities or UV efficiencies.
    }
    \label{fig:posterior}
\end{figure}

In order to visualize this, we show the results of varying both parameters over the entire range of $z=4-13$ in Fig.~\ref{fig:posterior} (see App.~\ref{app:bestfits} for the rest of astrophysical parameters).
In both models we find a consistent trend, where the parameter of interest ($\epsilon_{\star,\rm UV}$ or $\sUV$) has a fairly smooth, roughly constant behavior for $z\leq 10$, but abruptly rises at $z>10$ to explain the overabundance of JWST galaxies.
In the first case, we require very efficient UV emission with $\epsilon_{\star,\rm UV}\sim O(1)$, which may be due to feedback-free starbursts~\citep{Dekel:2023ddd}, or a top-heavy stellar initial mass function~\citep{Steinhardt_PopIII}.
In the second case, JWST data demand a very large stochasticity, $\sUV\approx 2$, which translates into a full dex of variation in the UV luminosity from galaxy to galaxy, much in excess of what is expected of mass-accretion histories~\citep{Mirocha:2020slz,Mason:2022tiy}.
Moreover, such large stochasticity may require coherent bursts of star formation lasting $\sim$ tens of Myrs, rather than short, noise-like variations, as spectral synthesis dampens the fluctuations in luminosity.
Let us now explore how to distinguish between these two scenarios.

\section{Clustering as a high-redshift model discriminator}
\label{sec:clusteringhiz}
Clustering measurements will provide information on the hosts of the high-$z$ galaxies. We show our predicted biases for both models discussed above in Fig.~\ref{fig:highz_uvlf}, along with the first JWST UVLFs at $z\sim 11$ and 12. 
The model with enhanced stochasticity predicts lower biases on average, as the more abundant lighter halos will dominate the bright end of the UVLF.
This is true even at magnitudes where the two models show similar UVLFs (e.g., $\MUV = -20$ to $-17$).
Thus, while different halo-galaxy connections may give rise to an identical one-point function (UVLF), the higher-order statistics (in our case the bias) will allow us to distinguish between models, even at $z\gtrsim 10$.

While the clustering of the first galaxies may hold the key to understanding their formation, measuring it is no small feat.
Studies using HST~\citep{Barone-Nugent:2014eqa,Lee:2005jha,Overzier:2006zf} and HSC~\citep{Harikane:2017lcw,Harikane_Goldrush_2021} have quantified clustering up to $z \sim 7$.
Extending these measurements to the $z\gtrsim 10$ regime is complicated due to the dearth of high-$z$ galaxies. There is reason, however, to be cautiously optimistic. The JWST-derived UVLFs are much larger than predicted, which means that if they hold to spectroscopic scrutiny we could expect significantly more targets at these $z$. Moreover, the expected biases are fairly large ($b_{\rm eff}\sim 8-10$, owing to the rarity of nonlinear structures at these high $z$), easing a detection of the clustering signal. Let us now perform a simple forecast.

We will focus on two survey configurations, the Cosmos-Web JWST survey, which we approximate as covering an area $\Omega=0.5\,\rm deg^2$, up to $m_{\rm UV} = 28.0$~\citep{Casey:2022amu}; and the proposed {\it Roman} HLS, covering $\Omega=2200\,\rm deg^2$, up to $m_{\rm UV} = 26.5$~\citep{Wang:2021oec}.
These represent two near- and mid-future possibilities to measure clustering at high $z$, though of course other data-sets, including deeper JWST and {\it Roman} surveys, will add further information.
For convenience, we will use a Fisher-matrix formalism as introduced in~\citet{Jungman:1995bz}:
\be
F_{bb} = f_{\rm sky} \sum_{\ell > \ell_{\rm min}} \left(\ell + \dfrac{1}{2} \right) \left(\dfrac{\partial C_\ell/\partial b_{\rm eff}}{C_\ell+N_\ell}\right)^2\ ,
\ee
with $\ell_{\rm min}=f_{\rm sky}^{-1/2}$ and $f_{\rm sky}=\Omega/(4\pi)$ the sky fraction covered by each survey.
We take the noise to be Poissonian, such that $N_\ell = \overline n_{2\mathrm{D}}^{-1}$, where
\be
\overline n_{2\mathrm{D}} = \phi_{\rm UV}^{\rm bin} \Delta \MUV \chi^2 \Delta \chi
\ee
is the surface density of galaxies in each bin, assuming an observational window centered at a comoving distance $\chi(z)$ with a width $\Delta \chi$.
As for the observed power spectrum, we take the Limber approximation~\citep{LoVerde:2008re}:
\be
C_{\ell} (z) = \dfrac{b_{\rm eff}^2}{(\Delta \chi)^2} \int_{\chi-\Delta \chi/2}^{\chi+\Delta \chi/2} \dfrac{\mathrm{d}\chi'}{\chi'^2} P_\mathrm{m}\left[(\ell+1/2)/\chi',z\right]\ ,
\ee
where we ignore the redshift evolution of $b_{\rm eff}$ and the matter power spectrum $P_\mathrm{m}$ in the integral.

Under these approximations, we can estimate the expected uncertainty in the bias as $\sigma(b_{\rm eff})= F_{bb}^{-1/2}$, which we show in Fig.~\ref{fig:highz_uvlf} at $z\sim 11$ and 12.
At both redshifts, we find that with Cosmos-Web we can expect a $\sim 20\%$ constraint on the bias over a $\Delta \MUV=2.0$ bin, potentially enough to distinguish between our two models and to pinpoint the mass of the halos that hosted such bright galaxies.
Moreover, the {\it Roman} HLS would reduce these errors to the $1-3\%$ level over several magnitude bins, providing a definitive test of our galaxy-formation models.

We note that this forecast was not designed to capture the detailed physics of galaxy clustering at high $z$, including covariances between data-points~\citep{Harikane_Goldrush_2021}, light-cone effects~\citep{Yung22_semianalytic_lightcone}, and the non-linear 1-halo term~\citep[][which would boost the signal]{Cooray:2002dia}.
Instead, it provides a proof-of-principle that
the biases we expect at $z\sim 10$ differ between galaxy-formation models, and are potentially measurable, given the enhancement in the expected number of galaxies from the JWST UVLFs.
We leave a detailed analysis for future work~\citep{Sabti_HSC_future}, including studying the mass dependence of $\sUV$, though we note that previous {\it Roman} forecasts in~\citet{Waters_2016_RomanWFIRST_forecast,LaPlante:2022nlp} have shown that the linear bias is potentially measurable up to $z\sim 13$.

\section{Discussion and Conclusions}
\label{sec:conclusions}

The first JWST-derived UV luminosity functions (UVLFs) at $z > 10$ point to a very active universe, with far more UV-bright galaxies than predicted by our models. 
The root cause of this discrepancy is, however, difficult to determine from the UVLFs alone.
This observable suffers from an intrinsic degeneracy between changes in the average halo-galaxy connection and its stochasticity, as illustrated in Fig.~\ref{fig:z4degeneracy}.
Here we have argued that galaxy-clustering measurements will allow us to break this degeneracy, providing an essential test of galaxy-formation models.

We have performed an analysis based on a flexible, semi-analytic model, which has been tested against HST data and hydrodynamical simulations in~\citet{Sabti:2021xvh}.
We calibrated our model to the $z \leq 10$ HST UVLFs and found that $\sUV$ (which accounts for stochasticity) and the star-formation efficiency $f_\star$ are indeed highly degenerate at each $z$.
However, we have shown that clustering measurements from~\citet[][using {\it Subaru} HSC]{Harikane_Goldrush_2021} tentatively point to a tight halo-galaxy connection at $z\sim 4-6$, with $\sUV\lesssim 0.5$.
This value is in line with the predicted variability in the assembly of dark-matter halos~\citep{Mirocha:2020slz},
as well as with previous studies using closest-neighbor clustering~\citep{Ren_2018_closest_neighbors} and field-to-field variations~\citep{Robertson_2010_biasCV}.

We have additionally applied our analysis to the first JWST UVLFs.
Theoretical work has argued for either an increased stochasticity ($\sUV$) or mean UV emission ($\epsilon_{\star,\rm UV}$) at $z\gtrsim10$ to explain the abundance of galaxy candidates in JWST.
We have shown that, while both solutions fit well the JWST UVLFs, the large-$\sUV$ scenario predicts consistently lower biases.
Further, these biases are potentially measurable at $z\gtrsim 10$ with the Cosmos-Web survey, and in the future by {\it Roman}, allowing us to distinguish between the two galaxy-formation scenarios.
Clustering can provide a more robust check than studying the bright-end of the UVLF alone, as the latter is susceptible to systematic effects such as AGN contamination and dust.

In summary, the launch of JWST has ignited a revolution in our understanding of the formation and evolution of the first galaxies.
Here we have argued that clustering measurements are critical to break degeneracies in galaxy models.
This will allow us to extract the maximum amount of information from JWST and future observatories like {\it Roman}, and will unveil the connection between the large-scale structure and the first luminous objects in our universe.

\subsection*{Acknowledgements}

We are very grateful to Mike Boylan-Kolchin, Caitlin Casey, Daniel Eisenstein, and Steve Finkelstein for helpful discussions, as well as to the anonymous referee for insightful comments.
JBM was supported by the UT Board of Visitors, and thanks the ITC at the Harvard-Smithsonian Center for Astrophysics for their hospitality during part of this work. 
JM was supported by an appointment to the NASA Postdoctoral Program at the Jet Propulsion Laboratory / California Institute of Technology, administered by Oak Ridge Associated Universities under contract with NASA.
SRF was supported by the National Science Foundation through awards AST-1812458 and AST-2205900 and by NASA through award 80NSSC22K0818. NS was supported by a Horizon Fellowship from Johns Hopkins University.

\subsection*{Data Availability}

The data underlying this article will be shared on reasonable request to the author.

{\it Software:} {\tt numpy}~\citep{numpy}, {\tt emcee}~\citep{Foreman-Mackey:2012any}, {\tt corner}~\citep{corner}, {\tt CLASS}~\citep{Blas:2011rf}, and {\tt Zeus21}~\citep{Munoz:2023kkg}.

\bibliographystyle{mnras}

\bibliography{jwst}

\begin{thebibliography}{}
\makeatletter
\relax
\def\mn@urlcharsother{\let\do\@makeother \do\$\do\&\do\#\do\^\do\_\do\%\do\~}
\def\mn@doi{\begingroup\mn@urlcharsother \@ifnextchar [ {\mn@doi@}
  {\mn@doi@[]}}
\def\mn@doi@[#1]#2{\def\@tempa{#1}\ifx\@tempa\@empty \href
  {http://dx.doi.org/#2} {doi:#2}\else \href {http://dx.doi.org/#2} {#1}\fi
  \endgroup}
\def\mn@eprint#1#2{\mn@eprint@#1:#2::\@nil}
\def\mn@eprint@arXiv#1{\href {http://arxiv.org/abs/#1} {{\tt arXiv:#1}}}
\def\mn@eprint@dblp#1{\href {http://dblp.uni-trier.de/rec/bibtex/#1.xml}
  {dblp:#1}}
\def\mn@eprint@#1:#2:#3:#4\@nil{\def\@tempa {#1}\def\@tempb {#2}\def\@tempc
  {#3}\ifx \@tempc \@empty \let \@tempc \@tempb \let \@tempb \@tempa \fi \ifx
  \@tempb \@empty \def\@tempb {arXiv}\fi \@ifundefined
  {mn@eprint@\@tempb}{\@tempb:\@tempc}{\expandafter \expandafter \csname
  mn@eprint@\@tempb\endcsname \expandafter{\@tempc}}}

\bibitem[\protect\citeauthoryear{{Adams} et~al.,}{{Adams}
  et~al.}{2023}]{Adams_Conselice_JWST_2023}
{Adams} N.~J.,  et~al., 2023, \mn@doi [\mnras] {10.1093/mnras/stac3347}, \href
  {https://ui.adsabs.harvard.edu/abs/2023MNRAS.518.4755A} {518, 4755}

\bibitem[\protect\citeauthoryear{Aghanim et~al.}{Aghanim
  et~al.}{2020}]{Planck:2018vyg}
Aghanim N.,  et~al., 2020, \mn@doi [Astron. Astrophys.]
  {10.1051/0004-6361/201833910}, 641, A6

\bibitem[\protect\citeauthoryear{{Arrabal Haro} et~al.,}{{Arrabal Haro}
  et~al.}{2023}]{Arrabal_CEERS_spectra}
{Arrabal Haro} P.,  et~al., 2023, \mn@doi [arXiv e-prints]
  {10.48550/arXiv.2303.15431}, \href
  {https://ui.adsabs.harvard.edu/abs/2023arXiv230315431A} {p. arXiv:2303.15431}

\bibitem[\protect\citeauthoryear{Atek et~al.}{Atek et~al.}{2015}]{Atek:2015axa}
Atek H.,  et~al., 2015, \mn@doi [Astrophys. J.] {10.1088/0004-637X/814/1/69},
  814, 69

\bibitem[\protect\citeauthoryear{Barone-Nugent et~al.}{Barone-Nugent
  et~al.}{2014}]{Barone-Nugent:2014eqa}
Barone-Nugent R.~L.,  et~al., 2014, \mn@doi [Astrophys. J.]
  {10.1088/0004-637X/793/1/17}, 793, 17

\bibitem[\protect\citeauthoryear{Behroozi, Conroy  \& Wechsler}{Behroozi
  et~al.}{2010}]{Behroozi:2010rx}
Behroozi P.~S.,  Conroy C.,   Wechsler R.~H.,  2010, \mn@doi [Astrophys. J.]
  {10.1088/0004-637X/717/1/379}, 717, 379

\bibitem[\protect\citeauthoryear{Berlind \& Weinberg}{Berlind \&
  Weinberg}{2002}]{Berlind:2001xk}
Berlind A.~A.,  Weinberg D.~H.,  2002, \mn@doi [Astrophys. J.]
  {10.1086/341469}, 575, 587

\bibitem[\protect\citeauthoryear{Blas, Lesgourgues  \& Tram}{Blas
  et~al.}{2011}]{Blas:2011rf}
Blas D.,  Lesgourgues J.,   Tram T.,  2011, \mn@doi [JCAP]
  {10.1088/1475-7516/2011/07/034}, 1107, 034

\bibitem[\protect\citeauthoryear{Bouwens et~al.,}{Bouwens
  et~al.}{2014}]{Bouwens:2013hxa}
Bouwens R.~J.,  et~al., 2014, \mn@doi [Astrophys. J.]
  {10.1088/0004-637X/793/2/115}, 793, 115

\bibitem[\protect\citeauthoryear{Bouwens et~al.}{Bouwens
  et~al.}{2015}]{Bouwens:2014fua}
Bouwens R.,  et~al., 2015, \mn@doi [Astrophys. J.]
  {10.1088/0004-637X/803/1/34}, 803, 34

\bibitem[\protect\citeauthoryear{{Bouwens}, {Oesch}, {Stefanon}, {Illingworth},
  {Labb{\'e}}  et~al.}{{Bouwens} et~al.}{2021}]{Bouwens_2021_UVLFs}
{Bouwens} R.~J.,  {Oesch} P.~A.,  {Stefanon} M.,  {Illingworth} G.,
  {Labb{\'e}} I.,   et~al., 2021, \mn@doi [Astrophys. J.]
  {10.3847/1538-3881/abf83e}, \href
  {https://ui.adsabs.harvard.edu/abs/2021AJ....162...47B} {162, 47}

\bibitem[\protect\citeauthoryear{{Bouwens}, {Illingworth}, {Oesch}, {Stefanon},
  {Naidu}, {van Leeuwen}  \& {Magee}}{{Bouwens} et~al.}{2023}]{Bouwens:2022gqg}
{Bouwens} R.,  {Illingworth} G.,  {Oesch} P.,  {Stefanon} M.,  {Naidu} R.,
  {van Leeuwen} I.,   {Magee} D.,  2023, \mn@doi [\mnras]
  {10.1093/mnras/stad1014}, \href
  {https://ui.adsabs.harvard.edu/abs/2023MNRAS.523.1009B} {523, 1009}

\bibitem[\protect\citeauthoryear{{Bromm}, {Coppi}  \& {Larson}}{{Bromm}
  et~al.}{2002}]{Bromm2001_firststars}
{Bromm} V.,  {Coppi} P.~S.,   {Larson} R.~B.,  2002, \mn@doi [\apj]
  {10.1086/323947}, \href
  {https://ui.adsabs.harvard.edu/abs/2002ApJ...564...23B} {564, 23}

\bibitem[\protect\citeauthoryear{Casey et~al.}{Casey
  et~al.}{2022}]{Casey:2022amu}
Casey C.~M.,  et~al., 2022

\bibitem[\protect\citeauthoryear{{Castellano}, {Fontana}, {Treu}, {Santini},
  {Merlin}  et~al.}{{Castellano} et~al.}{2022}]{Castellano_GLASS_hiz}
{Castellano} M.,  {Fontana} A.,  {Treu} T.,  {Santini} P.,  {Merlin} E.,
  et~al., 2022, \mn@doi [Astrophys. J. Lett.] {10.3847/2041-8213/ac94d0}, \href
  {https://ui.adsabs.harvard.edu/abs/2022ApJ...938L..15C} {938, L15}

\bibitem[\protect\citeauthoryear{{Ceverino}, {Glover}  \& {Klessen}}{{Ceverino}
  et~al.}{2017}]{Ceverino_FirstLight_UVLF}
{Ceverino} D.,  {Glover} S. C.~O.,   {Klessen} R.~S.,  2017, \mn@doi [\mnras]
  {10.1093/mnras/stx1386}, \href
  {https://ui.adsabs.harvard.edu/abs/2017MNRAS.470.2791C} {470, 2791}

\bibitem[\protect\citeauthoryear{Cooray \& Sheth}{Cooray \&
  Sheth}{2002}]{Cooray:2002dia}
Cooray A.,  Sheth R.~K.,  2002, \mn@doi [Phys. Rept.]
  {10.1016/S0370-1573(02)00276-4}, 372, 1

\bibitem[\protect\citeauthoryear{Corasaniti, Agarwal, Marsh  \& Das}{Corasaniti
  et~al.}{2017}]{Corasaniti:2016epp}
Corasaniti P.,  Agarwal S.,  Marsh D.,   Das S.,  2017, \mn@doi [Phys. Rev. D]
  {10.1103/PhysRevD.95.083512}, 95, 083512

\bibitem[\protect\citeauthoryear{{Curtis-Lake} et~al.,}{{Curtis-Lake}
  et~al.}{2022}]{Curtis-Lake2022:JWST}
{Curtis-Lake} E.,  et~al., 2022, \mn@doi [arXiv e-prints]
  {10.48550/arXiv.2212.04568}, \href
  {https://ui.adsabs.harvard.edu/abs/2022arXiv221204568C} {p. arXiv:2212.04568}

\bibitem[\protect\citeauthoryear{{Dekel}, {Sarkar}, {Birnboim}, {Mandelker}  \&
  {Li}}{{Dekel} et~al.}{2023}]{Dekel:2023ddd}
{Dekel} A.,  {Sarkar} K.~C.,  {Birnboim} Y.,  {Mandelker} N.,   {Li} Z.,  2023,
  \mn@doi [\mnras] {10.1093/mnras/stad1557}, \href
  {https://ui.adsabs.harvard.edu/abs/2023MNRAS.523.3201D} {523, 3201}

\bibitem[\protect\citeauthoryear{Desjacques, Jeong  \& Schmidt}{Desjacques
  et~al.}{2018}]{Desjacques:2016bnm}
Desjacques V.,  Jeong D.,   Schmidt F.,  2018, \mn@doi [Phys. Rept.]
  {10.1016/j.physrep.2017.12.002}, 733, 1

\bibitem[\protect\citeauthoryear{{Donnan}, {McLeod}, {McLure}, {Dunlop},
  {Carnall}  et~al.}{{Donnan} et~al.}{2023}]{Donnan_UVLFs_2023}
{Donnan} C.~T.,  {McLeod} D.~J.,  {McLure} R.~J.,  {Dunlop} J.~S.,  {Carnall}
  A.~C.,   et~al., 2023, \mn@doi [Mon. Notices Royal Astron. Soc.]
  {10.1093/mnras/stad471}, \href
  {https://ui.adsabs.harvard.edu/abs/2023MNRAS.520.4554D} {520, 4554}

\bibitem[\protect\citeauthoryear{{Eisenstein} et~al.,}{{Eisenstein}
  et~al.}{2023}]{Eisenstein_JADES}
{Eisenstein} D.~J.,  et~al., 2023, \mn@doi [arXiv e-prints]
  {10.48550/arXiv.2306.02465}, \href
  {https://ui.adsabs.harvard.edu/abs/2023arXiv230602465E} {p. arXiv:2306.02465}

\bibitem[\protect\citeauthoryear{Fakhouri, Ma  \& Boylan-Kolchin}{Fakhouri
  et~al.}{2010}]{Fakhouri:2010st}
Fakhouri O.,  Ma C.-P.,   Boylan-Kolchin M.,  2010, \mn@doi [Mon. Not. Roy.
  Astron. Soc.] {10.1111/j.1365-2966.2010.16859.x}, 406, 2267

\bibitem[\protect\citeauthoryear{{Faucher-Gigu{\`e}re}}{{Faucher-Gigu{\`e}re}}{2018}]{Faucher_2018_feedback}
{Faucher-Gigu{\`e}re} C.-A.,  2018, \mn@doi [\mnras] {10.1093/mnras/stx2595},
  \href {https://ui.adsabs.harvard.edu/abs/2018MNRAS.473.3717F} {473, 3717}

\bibitem[\protect\citeauthoryear{{Ferrara}, {Pallottini}  \& {Dayal}}{{Ferrara}
  et~al.}{2022}]{Ferrara2022}
{Ferrara} A.,  {Pallottini} A.,   {Dayal} P.,  2022, arXiv e-prints, \href
  {https://ui.adsabs.harvard.edu/abs/2022arXiv220800720F} {p. arXiv:2208.00720}

\bibitem[\protect\citeauthoryear{{Finkelstein}}{{Finkelstein}}{2016}]{Finkelstein_2015_review}
{Finkelstein} S.~L.,  2016, \mn@doi [\pasa] {10.1017/pasa.2016.26}, \href
  {https://ui.adsabs.harvard.edu/abs/2016PASA...33...37F} {33, e037}

\bibitem[\protect\citeauthoryear{{Finkelstein} \& {Bagley}}{{Finkelstein} \&
  {Bagley}}{2022}]{Finkelstein_Bagley_22}
{Finkelstein} S.~L.,  {Bagley} M.~B.,  2022, \mn@doi [\apj]
  {10.3847/1538-4357/ac89eb}, \href
  {https://ui.adsabs.harvard.edu/abs/2022ApJ...938...25F} {938, 25}

\bibitem[\protect\citeauthoryear{{Finkelstein} et~al.,}{{Finkelstein}
  et~al.}{2022a}]{Finkelstein_CEERS}
{Finkelstein} S.~L.,  et~al., 2022a, \mn@doi [arXiv e-prints]
  {10.48550/arXiv.2211.05792}, \href
  {https://ui.adsabs.harvard.edu/abs/2022arXiv221105792F} {p. arXiv:2211.05792}

\bibitem[\protect\citeauthoryear{{Finkelstein} et~al.,}{{Finkelstein}
  et~al.}{2022b}]{Finkelstein_2022_Maisies}
{Finkelstein} S.~L.,  et~al., 2022b, \mn@doi [\apjl]
  {10.3847/2041-8213/ac966e}, \href
  {https://ui.adsabs.harvard.edu/abs/2022ApJ...940L..55F} {940, L55}

\bibitem[\protect\citeauthoryear{Foreman-Mackey}{Foreman-Mackey}{2016}]{corner}
Foreman-Mackey D.,  2016, \mn@doi [The Journal of Open Source Software]
  {10.21105/joss.00024}, 1, 24

\bibitem[\protect\citeauthoryear{Foreman-Mackey, Hogg, Lang  \&
  Goodman}{Foreman-Mackey et~al.}{2013}]{Foreman-Mackey:2012any}
Foreman-Mackey D.,  Hogg D.~W.,  Lang D.,   Goodman J.,  2013, \mn@doi [Publ.
  Astron. Soc. Pac.] {10.1086/670067}, 125, 306

\bibitem[\protect\citeauthoryear{{Fujimoto} et~al.,}{{Fujimoto}
  et~al.}{2023}]{Fujimoto:2023orx}
{Fujimoto} S.,  et~al., 2023, \mn@doi [\apjl] {10.3847/2041-8213/acd2d9}, \href
  {https://ui.adsabs.harvard.edu/abs/2023ApJ...949L..25F} {949, L25}

\bibitem[\protect\citeauthoryear{{Furlanetto} \& {Mirocha}}{{Furlanetto} \&
  {Mirocha}}{2022a}]{Furlanetto_purity_JWST}
{Furlanetto} S.~R.,  {Mirocha} J.,  2022a, \mn@doi [arXiv e-prints]
  {10.48550/arXiv.2208.12828}, \href
  {https://ui.adsabs.harvard.edu/abs/2022arXiv220812828F} {p. arXiv:2208.12828}

\bibitem[\protect\citeauthoryear{{Furlanetto} \& {Mirocha}}{{Furlanetto} \&
  {Mirocha}}{2022b}]{Furlanetto_Mirocha_21_bursty}
{Furlanetto} S.~R.,  {Mirocha} J.,  2022b, \mn@doi [\mnras]
  {10.1093/mnras/stac310}, \href
  {https://ui.adsabs.harvard.edu/abs/2022MNRAS.511.3895F} {511, 3895}

\bibitem[\protect\citeauthoryear{{Furlanetto}, {Mirocha}, {Mebane}  \&
  {Sun}}{{Furlanetto} et~al.}{2017}]{Furlanetto2016:feedback}
{Furlanetto} S.~R.,  {Mirocha} J.,  {Mebane} R.~H.,   {Sun} G.,  2017, \mn@doi
  [\mnras] {10.1093/mnras/stx2132}, \href
  {https://ui.adsabs.harvard.edu/abs/2017MNRAS.472.1576F} {472, 1576}

\bibitem[\protect\citeauthoryear{Giavalisco \& Dickinson}{Giavalisco \&
  Dickinson}{2001}]{Giavalisco:2000xs}
Giavalisco M.,  Dickinson M.,  2001, \mn@doi [Astrophys. J.] {10.1086/319715},
  550, 177

\bibitem[\protect\citeauthoryear{Hadzhiyska, Liu, Somerville, Gabrielpillai,
  Bose, Eisenstein  \& Hernquist}{Hadzhiyska et~al.}{2021}]{Hadzhiyska:2021kmt}
Hadzhiyska B.,  Liu S.,  Somerville R.~S.,  Gabrielpillai A.,  Bose S.,
  Eisenstein D.,   Hernquist L.,  2021, \mn@doi [Mon. Not. Roy. Astron. Soc.]
  {10.1093/mnras/stab2564}, 508, 698

\bibitem[\protect\citeauthoryear{{Harikane} et~al.,}{{Harikane}
  et~al.}{2018}]{Harikane:2017lcw}
{Harikane} Y.,  et~al., 2018, \mn@doi [\pasj] {10.1093/pasj/psx097}, \href
  {https://ui.adsabs.harvard.edu/abs/2018PASJ...70S..11H} {70, S11}

\bibitem[\protect\citeauthoryear{{Harikane}, {Ono}, {Ouchi}, {Liu}, {Sawicki}
  et~al.}{{Harikane} et~al.}{2022}]{Harikane_Goldrush_2021}
{Harikane} Y.,  {Ono} Y.,  {Ouchi} M.,  {Liu} C.,  {Sawicki} M.,   et~al.,
  2022, \mn@doi [Astrophys. J., Suppl. Ser.] {10.3847/1538-4365/ac3dfc}, \href
  {https://ui.adsabs.harvard.edu/abs/2022ApJS..259...20H} {259, 20}

\bibitem[\protect\citeauthoryear{{Harikane}, {Nakajima}, {Ouchi}, {Umeda},
  {Isobe}, {Ono}, {Xu}  \& {Zhang}}{{Harikane}
  et~al.}{2023a}]{Harikane_specz_UVLF_JWST}
{Harikane} Y.,  {Nakajima} K.,  {Ouchi} M.,  {Umeda} H.,  {Isobe} Y.,  {Ono}
  Y.,  {Xu} Y.,   {Zhang} Y.,  2023a, \mn@doi [arXiv e-prints]
  {10.48550/arXiv.2304.06658}, \href
  {https://ui.adsabs.harvard.edu/abs/2023arXiv230406658H} {p. arXiv:2304.06658}

\bibitem[\protect\citeauthoryear{{Harikane}, {Ouchi}, {Oguri}, {Ono},
  {Nakajima}  et~al.}{{Harikane} et~al.}{2023b}]{Harikane_UVLFs}
{Harikane} Y.,  {Ouchi} M.,  {Oguri} M.,  {Ono} Y.,  {Nakajima} K.,   et~al.,
  2023b, \mn@doi [Astrophys. J., Suppl. Ser.] {10.3847/1538-4365/acaaa9}, \href
  {https://ui.adsabs.harvard.edu/abs/2023ApJS..265....5H} {265, 5}

\bibitem[\protect\citeauthoryear{{Inayoshi}, {Harikane}, {Inoue}, {Li}  \&
  {Ho}}{{Inayoshi} et~al.}{2022}]{Inayoshi_JWST_2023}
{Inayoshi} K.,  {Harikane} Y.,  {Inoue} A.~K.,  {Li} W.,   {Ho} L.~C.,  2022,
  \mn@doi [\apjl] {10.3847/2041-8213/ac9310}, \href
  {https://ui.adsabs.harvard.edu/abs/2022ApJ...938L..10I} {938, L10}

\bibitem[\protect\citeauthoryear{Jungman, Kamionkowski, Kosowsky  \&
  Spergel}{Jungman et~al.}{1996}]{Jungman:1995bz}
Jungman G.,  Kamionkowski M.,  Kosowsky A.,   Spergel D.~N.,  1996, \mn@doi
  [Phys. Rev. D] {10.1103/PhysRevD.54.1332}, 54, 1332

\bibitem[\protect\citeauthoryear{{Kauffmann} et~al.,}{{Kauffmann}
  et~al.}{2022}]{Cosmos2020_Kauffman}
{Kauffmann} O.~B.,  et~al., 2022, \mn@doi [\aap] {10.1051/0004-6361/202243088},
  \href {https://ui.adsabs.harvard.edu/abs/2022A&A...667A..65K} {667, A65}

\bibitem[\protect\citeauthoryear{{Kocevski} et~al.,}{{Kocevski}
  et~al.}{2023}]{Kocevski_CEERS_spectra_AGN}
{Kocevski} D.~D.,  et~al., 2023, \mn@doi [\apjl] {10.3847/2041-8213/acad00},
  \href {https://ui.adsabs.harvard.edu/abs/2023ApJ...946L..14K} {946, L14}

\bibitem[\protect\citeauthoryear{Kravtsov, Berlind, Wechsler, Klypin,
  Gottloeber, Allgood  \& Primack}{Kravtsov et~al.}{2004}]{Kravtsov:2003sg}
Kravtsov A.~V.,  Berlind A.~A.,  Wechsler R.~H.,  Klypin A.~A.,  Gottloeber S.,
   Allgood B.,   Primack J.~R.,  2004, \mn@doi [Astrophys. J.]
  {10.1086/420959}, 609, 35

\bibitem[\protect\citeauthoryear{La~Plante, Mirocha, Gorce, Lidz  \&
  Parsons}{La~Plante et~al.}{2023}]{LaPlante:2022nlp}
La~Plante P.,  Mirocha J.,  Gorce A.,  Lidz A.,   Parsons A.,  2023, \mn@doi
  [Astrophys. J.] {10.3847/1538-4357/acaeb0}, 944, 59

\bibitem[\protect\citeauthoryear{Lee, Giavalisco, Gnedin, Somerville, Ferguson,
  Dickinson  \& Ouchi}{Lee et~al.}{2006}]{Lee:2005jha}
Lee K.,  Giavalisco M.,  Gnedin O.~Y.,  Somerville R.,  Ferguson H.,  Dickinson
  M.,   Ouchi M.,  2006, \mn@doi [Astrophys. J.] {10.1086/500387}, 642, 63

\bibitem[\protect\citeauthoryear{{Leung} et~al.,}{{Leung}
  et~al.}{2023}]{Leung_NGdeep_2023}
{Leung} G. C.~K.,  et~al., 2023, \mn@doi [arXiv e-prints]
  {10.48550/arXiv.2306.06244}, \href
  {https://ui.adsabs.harvard.edu/abs/2023arXiv230606244L} {p. arXiv:2306.06244}

\bibitem[\protect\citeauthoryear{Livermore, Finkelstein  \& Lotz}{Livermore
  et~al.}{2017}]{Livermore:2016mbs}
Livermore R.,  Finkelstein S.,   Lotz J.,  2017, \mn@doi [Astrophys. J.]
  {10.3847/1538-4357/835/2/113}, 835, 113

\bibitem[\protect\citeauthoryear{LoVerde \& Afshordi}{LoVerde \&
  Afshordi}{2008}]{LoVerde:2008re}
LoVerde M.,  Afshordi N.,  2008, \mn@doi [Phys. Rev. D]
  {10.1103/PhysRevD.78.123506}, 78, 123506

\bibitem[\protect\citeauthoryear{Madau \& Dickinson}{Madau \&
  Dickinson}{2014}]{Madau:2014bja}
Madau P.,  Dickinson M.,  2014, \mn@doi [Ann. Rev. Astron. Astrophys.]
  {10.1146/annurev-astro-081811-125615}, 52, 415

\bibitem[\protect\citeauthoryear{{Maiolino} et~al.,}{{Maiolino}
  et~al.}{2023}]{Maiolino:2023zdu}
{Maiolino} R.,  et~al., 2023, \mn@doi [arXiv e-prints]
  {10.48550/arXiv.2305.12492}, \href
  {https://ui.adsabs.harvard.edu/abs/2023arXiv230512492M} {p. arXiv:2305.12492}

\bibitem[\protect\citeauthoryear{Mason, Naidu, Tacchella  \& Leja}{Mason
  et~al.}{2019}]{Mason:2019oeg}
Mason C.~A.,  Naidu R.~P.,  Tacchella S.,   Leja J.,  2019, \mn@doi [Mon. Not.
  Roy. Astron. Soc.] {10.1093/mnras/stz2291}, 489, 2669

\bibitem[\protect\citeauthoryear{Mason, Trenti  \& Treu}{Mason
  et~al.}{2023}]{Mason:2022tiy}
Mason C.~A.,  Trenti M.,   Treu T.,  2023, \mn@doi [Mon. Not. Roy. Astron.
  Soc.] {10.1093/mnras/stad035}, 521, 497

\bibitem[\protect\citeauthoryear{Menci, Grazian, Lamastra, Calura, Castellano
  et~al.}{Menci et~al.}{2018}]{Menci:2018lis}
Menci N.,  Grazian A.,  Lamastra A.,  Calura F.,  Castellano M.,   et~al.,
  2018, \mn@doi [Astrophys. J.] {10.3847/1538-4357/aaa773}, 854, 1

\bibitem[\protect\citeauthoryear{Meurer, Heckman  \& Calzetti}{Meurer
  et~al.}{1999}]{Meurer:1999jj}
Meurer G.~R.,  Heckman T.~M.,   Calzetti D.,  1999, \mn@doi [Astrophys. J.]
  {10.1086/307523}, 521, 64

\bibitem[\protect\citeauthoryear{{Mirocha}}{{Mirocha}}{2020}]{Mirocha:2020bias}
{Mirocha} J.,  2020, \mn@doi [\mnras] {10.1093/mnras/staa3150}, \href
  {https://ui.adsabs.harvard.edu/abs/2020MNRAS.499.4534M} {499, 4534}

\bibitem[\protect\citeauthoryear{{Mirocha} \& {Furlanetto}}{{Mirocha} \&
  {Furlanetto}}{2023}]{Mirocha_UVLFs2023}
{Mirocha} J.,  {Furlanetto} S.~R.,  2023, \mn@doi [Mon. Notices Royal Astron.
  Soc.] {10.1093/mnras/stac3578}, \href
  {https://ui.adsabs.harvard.edu/abs/2023MNRAS.519..843M} {519, 843}

\bibitem[\protect\citeauthoryear{{Mirocha}, {Furlanetto}  \& {Sun}}{{Mirocha}
  et~al.}{2017}]{Mirocha_UVLFs_2017}
{Mirocha} J.,  {Furlanetto} S.~R.,   {Sun} G.,  2017, \mn@doi [\mnras]
  {10.1093/mnras/stw2412}, \href
  {https://ui.adsabs.harvard.edu/abs/2017MNRAS.464.1365M} {464, 1365}

\bibitem[\protect\citeauthoryear{Mirocha, La~Plante  \& Liu}{Mirocha
  et~al.}{2021}]{Mirocha:2020slz}
Mirocha J.,  La~Plante P.,   Liu A.,  2021, \mn@doi [Mon. Not. Roy. Astron.
  Soc.] {10.1093/mnras/stab1871}, 507, 3872

\bibitem[\protect\citeauthoryear{{Moster}, {Somerville}, {Maulbetsch}, {van den
  Bosch}, {Macci{\`o}}, {Naab}  \& {Oser}}{{Moster}
  et~al.}{2010}]{Moster:2009fk}
{Moster} B.~P.,  {Somerville} R.~S.,  {Maulbetsch} C.,  {van den Bosch} F.~C.,
  {Macci{\`o}} A.~V.,  {Naab} T.,   {Oser} L.,  2010, \mn@doi [\apj]
  {10.1088/0004-637X/710/2/903}, \href
  {https://ui.adsabs.harvard.edu/abs/2010ApJ...710..903M} {710, 903}

\bibitem[\protect\citeauthoryear{Moster, Naab  \& White}{Moster
  et~al.}{2013}]{Moster:2012fv}
Moster B.~P.,  Naab T.,   White S. D.~M.,  2013, \mn@doi [Mon. Not. Roy.
  Astron. Soc.] {10.1093/mnras/sts261}, 428, 3121

\bibitem[\protect\citeauthoryear{{Mu{\~n}oz}}{{Mu{\~n}oz}}{2023}]{Munoz:2023kkg}
{Mu{\~n}oz} J.~B.,  2023, \mn@doi [\mnras] {10.1093/mnras/stad1512}, \href
  {https://ui.adsabs.harvard.edu/abs/2023MNRAS.523.2587M} {523, 2587}

\bibitem[\protect\citeauthoryear{{Naidu} et~al.,}{{Naidu}
  et~al.}{2022}]{Naidu2022}
{Naidu} R.~P.,  et~al., 2022, \mn@doi [\apjl] {10.3847/2041-8213/ac9b22}, \href
  {https://ui.adsabs.harvard.edu/abs/2022ApJ...940L..14N} {940, L14}

\bibitem[\protect\citeauthoryear{Neistein \& van~den Bosch}{Neistein \& van~den
  Bosch}{2006}]{Neistein:2006ak}
Neistein E.,  van~den Bosch F.~C.,  2006, \mn@doi [Mon. Not. Roy. Astron. Soc.]
  {10.1111/j.1365-2966.2006.10918.x}, 372, 933

\bibitem[\protect\citeauthoryear{{Oesch}, {Brammer}, {van Dokkum},
  {Illingworth}, {Bouwens}  et~al.}{{Oesch} et~al.}{2016}]{Gnz11_Oesch}
{Oesch} P.~A.,  {Brammer} G.,  {van Dokkum} P.~G.,  {Illingworth} G.~D.,
  {Bouwens} R.~J.,   et~al., 2016, \mn@doi [Astrophys. J.]
  {10.3847/0004-637X/819/2/129}, \href
  {https://ui.adsabs.harvard.edu/abs/2016ApJ...819..129O} {819, 129}

\bibitem[\protect\citeauthoryear{{Oke} \& {Gunn}}{{Oke} \&
  {Gunn}}{1983}]{Oke_Gunn_AB_mags}
{Oke} J.~B.,  {Gunn} J.~E.,  1983, \mn@doi [\apj] {10.1086/160817}, \href
  {https://ui.adsabs.harvard.edu/abs/1983ApJ...266..713O} {266, 713}

\bibitem[\protect\citeauthoryear{Overzier, Bouwens, Illingworth  \&
  Franx}{Overzier et~al.}{2006}]{Overzier:2006zf}
Overzier R.~A.,  Bouwens R.~J.,  Illingworth G.~D.,   Franx M.,  2006, \mn@doi
  [Astrophys. J. Lett.] {10.1086/507678}, 648, L5

\bibitem[\protect\citeauthoryear{{Padmanabhan} \& {Loeb}}{{Padmanabhan} \&
  {Loeb}}{2023}]{Padmanabhan:2023esp}
{Padmanabhan} H.,  {Loeb} A.,  2023, \mn@doi [arXiv e-prints]
  {10.48550/arXiv.2306.04684}, \href
  {https://ui.adsabs.harvard.edu/abs/2023arXiv230604684P} {p. arXiv:2306.04684}

\bibitem[\protect\citeauthoryear{Park, Mesinger, Greig  \& Gillet}{Park
  et~al.}{2019}]{Park:2018ljd}
Park J.,  Mesinger A.,  Greig B.,   Gillet N.,  2019, \mn@doi [Mon. Not. Roy.
  Astron. Soc.] {10.1093/mnras/stz032}, 484, 933

\bibitem[\protect\citeauthoryear{P\'erez-Gonz\'alez et~al.}{P\'erez-Gonz\'alez
  et~al.}{2023}]{Perez-Gonzalez:2023wta}
P\'erez-Gonz\'alez P.~G.,  et~al., 2023

\bibitem[\protect\citeauthoryear{{Ren}, {Trenti}  \& {Mutch}}{{Ren}
  et~al.}{2018}]{Ren_2018_closest_neighbors}
{Ren} K.,  {Trenti} M.,   {Mutch} S.~J.,  2018, \mn@doi [\apj]
  {10.3847/1538-4357/aab094}, \href
  {https://ui.adsabs.harvard.edu/abs/2018ApJ...856...81R} {856, 81}

\bibitem[\protect\citeauthoryear{Ren, Trenti  \& Mason}{Ren
  et~al.}{2019}]{Ren_2019_scatter}
Ren K.,  Trenti M.,   Mason C.~A.,  2019, \mn@doi [The Astrophysical Journal]
  {10.3847/1538-4357/ab2117}, 878, 114

\bibitem[\protect\citeauthoryear{{Robertson}}{{Robertson}}{2010}]{Robertson_2010_biasCV}
{Robertson} B.~E.,  2010, \mn@doi [\apjl] {10.1088/2041-8205/716/2/L229}, \href
  {https://ui.adsabs.harvard.edu/abs/2010ApJ...716L.229R} {716, L229}

\bibitem[\protect\citeauthoryear{Rodriguez-Puebla, Behroozi, Primack, Klypin,
  Lee  \& Hellinger}{Rodriguez-Puebla et~al.}{2016}]{Rodriguez-Puebla:2016ofw}
Rodriguez-Puebla A.,  Behroozi P.,  Primack J.,  Klypin A.,  Lee C.,
  Hellinger D.,  2016, \mn@doi [Mon. Not. Roy. Astron. Soc.]
  {10.1093/mnras/stw1705}, 462, 893

\bibitem[\protect\citeauthoryear{{Rudakovskyi}, {Mesinger}, {Savchenko}  \&
  {Gillet}}{{Rudakovskyi} et~al.}{2021}]{Rudakovskyi:2021jyf}
{Rudakovskyi} A.,  {Mesinger} A.,  {Savchenko} D.,   {Gillet} N.,  2021,
  \mn@doi [\mnras] {10.1093/mnras/stab2333}, \href
  {https://ui.adsabs.harvard.edu/abs/2021MNRAS.507.3046R} {507, 3046}

\bibitem[\protect\citeauthoryear{Sabti, Mu\~noz  \& Blas}{Sabti
  et~al.}{2021}]{Sabti:2020ser}
Sabti N.,  Mu\~noz J.~B.,   Blas D.,  2021, \mn@doi [JCAP]
  {10.1088/1475-7516/2021/01/010}, 01, 010

\bibitem[\protect\citeauthoryear{Sabti, Mu\~noz  \& Blas}{Sabti
  et~al.}{2022a}]{Sabti:2021xvh}
Sabti N.,  Mu\~noz J.~B.,   Blas D.,  2022a, \mn@doi [Phys. Rev. D]
  {10.1103/PhysRevD.105.043518}, 105, 043518

\bibitem[\protect\citeauthoryear{Sabti, Mu\~noz  \& Blas}{Sabti
  et~al.}{2022b}]{Sabti:2021unj}
Sabti N.,  Mu\~noz J.~B.,   Blas D.,  2022b, \mn@doi [Astrophys. J. Lett.]
  {10.3847/2041-8213/ac5e9c}, 928, L20

\bibitem[\protect\citeauthoryear{Sabti, Mu\~noz  et~al.}{Sabti
  et~al.}{2023a}]{Sabti_HSC_future}
Sabti N.,  Mu\~noz J.~B.,   et~al., 2023a, In prep

\bibitem[\protect\citeauthoryear{{Sabti}, {Mu{\~n}oz}  \&
  {Kamionkowski}}{{Sabti} et~al.}{2023b}]{Sabti:2023xwo}
{Sabti} N.,  {Mu{\~n}oz} J.~B.,   {Kamionkowski} M.,  2023b, \mn@doi [arXiv
  e-prints] {10.48550/arXiv.2305.07049}, \href
  {https://ui.adsabs.harvard.edu/abs/2023arXiv230507049S} {p. arXiv:2305.07049}

\bibitem[\protect\citeauthoryear{Schneider, Giri  \& Mirocha}{Schneider
  et~al.}{2021}]{Schneider:2020xmf}
Schneider A.,  Giri S.~K.,   Mirocha J.,  2021, \mn@doi [Phys. Rev. D]
  {10.1103/PhysRevD.103.083025}, 103, 083025

\bibitem[\protect\citeauthoryear{{Shen}, {Vogelsberger}, {Boylan-Kolchin},
  {Tacchella}  \& {Kannan}}{{Shen} et~al.}{2023}]{Shen:2023cva}
{Shen} X.,  {Vogelsberger} M.,  {Boylan-Kolchin} M.,  {Tacchella} S.,
  {Kannan} R.,  2023, \mn@doi [arXiv e-prints] {10.48550/arXiv.2305.05679},
  \href {https://ui.adsabs.harvard.edu/abs/2023arXiv230505679S} {p.
  arXiv:2305.05679}

\bibitem[\protect\citeauthoryear{Sheth \& Tormen}{Sheth \&
  Tormen}{2002}]{Sheth:2001dp}
Sheth R.~K.,  Tormen G.,  2002, \mn@doi [Mon. Not. Roy. Astron. Soc.]
  {10.1046/j.1365-8711.2002.04950.x}, 329, 61

\bibitem[\protect\citeauthoryear{{Spergel} et~al.,}{{Spergel}
  et~al.}{2015}]{Spergel:2015sza}
{Spergel} D.,  et~al., 2015, \mn@doi [arXiv e-prints]
  {10.48550/arXiv.1503.03757}, \href
  {https://ui.adsabs.harvard.edu/abs/2015arXiv150303757S} {p. arXiv:1503.03757}

\bibitem[\protect\citeauthoryear{{Steinhardt}, {Sneppen}, {Mostafa}, {Hensley},
  {Jermyn}  et~al.}{{Steinhardt} et~al.}{2022}]{Steinhardt_PopIII}
{Steinhardt} C.~L.,  {Sneppen} A.,  {Mostafa} B.,  {Hensley} H.,  {Jermyn}
  A.~S.,   et~al., 2022, \mn@doi [Astrophys. J.] {10.3847/1538-4357/ac62d6},
  \href {https://ui.adsabs.harvard.edu/abs/2022ApJ...931...58S} {931, 58}

\bibitem[\protect\citeauthoryear{Tacchella, Bose, Conroy, Eisenstein  \&
  Johnson}{Tacchella et~al.}{2018}]{Tacchella:2018qny}
Tacchella S.,  Bose S.,  Conroy C.,  Eisenstein D.~J.,   Johnson B.~D.,  2018,
  \mn@doi [Astrophys. J.] {10.3847/1538-4357/aae8e0}, 868, 92

\bibitem[\protect\citeauthoryear{Tinker, Robertson, Kravtsov, Klypin, Warren,
  Yepes  \& Gottlober}{Tinker et~al.}{2010}]{Tinker:2010my}
Tinker J.~L.,  Robertson B.~E.,  Kravtsov A.~V.,  Klypin A.,  Warren M.~S.,
  Yepes G.,   Gottlober S.,  2010, \mn@doi [Astrophys. J.]
  {10.1088/0004-637X/724/2/878}, 724, 878

\bibitem[\protect\citeauthoryear{Trac, Cen  \& Mansfield}{Trac
  et~al.}{2015}]{Trac:2015rva}
Trac H.,  Cen R.,   Mansfield P.,  2015, \mn@doi [Astrophys. J.]
  {10.1088/0004-637X/813/1/54}, 813, 54

\bibitem[\protect\citeauthoryear{Trapp, Furlanetto  \& Yang}{Trapp
  et~al.}{2022}]{Trapp:2021ufe}
Trapp A.~C.,  Furlanetto S.~R.,   Yang J.,  2022, \mn@doi [Mon. Not. Roy.
  Astron. Soc.] {10.1093/mnras/stab3801}, 510, 4844

\bibitem[\protect\citeauthoryear{Trenti, Stiavelli, Bouwens, Oesch, Shull,
  Illingworth, Bradley  \& Carollo}{Trenti et~al.}{2010}]{Trenti:2010sz}
Trenti M.,  Stiavelli M.,  Bouwens R.~J.,  Oesch P.,  Shull J.~M.,  Illingworth
  G.~D.,  Bradley L.~D.,   Carollo C.~M.,  2010, \mn@doi [Astrophys. J. Lett.]
  {10.1088/2041-8205/714/2/L202}, 714, L202

\bibitem[\protect\citeauthoryear{Treu et~al.}{Treu et~al.}{2022}]{Treu:2022iti}
Treu T.,  et~al., 2022, \mn@doi [Astrophys. J.] {10.3847/1538-4357/ac8158},
  935, 110

\bibitem[\protect\citeauthoryear{Wang et~al.}{Wang et~al.}{2022}]{Wang:2021oec}
Wang Y.,  et~al., 2022, \mn@doi [Astrophys. J.] {10.3847/1538-4357/ac4973},
  928, 1

\bibitem[\protect\citeauthoryear{{Waters}, {Di Matteo}, {Feng}, {Wilkins}  \&
  {Croft}}{{Waters} et~al.}{2016}]{Waters_2016_RomanWFIRST_forecast}
{Waters} D.,  {Di Matteo} T.,  {Feng} Y.,  {Wilkins} S.~M.,   {Croft} R. A.~C.,
   2016, \mn@doi [\mnras] {10.1093/mnras/stw2000}, \href
  {https://ui.adsabs.harvard.edu/abs/2016MNRAS.463.3520W} {463, 3520}

\bibitem[\protect\citeauthoryear{Wechsler \& Tinker}{Wechsler \&
  Tinker}{2018}]{Wechsler:2018pic}
Wechsler R.~H.,  Tinker J.~L.,  2018, \mn@doi [Ann. Rev. Astron. Astrophys.]
  {10.1146/annurev-astro-081817-051756}, 56, 435

\bibitem[\protect\citeauthoryear{Xavier, Abdalla  \& Joachimi}{Xavier
  et~al.}{2016}]{Xavier:2016elr}
Xavier H.~S.,  Abdalla F.~B.,   Joachimi B.,  2016, \mn@doi [Mon. Not. Roy.
  Astron. Soc.] {10.1093/mnras/stw874}, 459, 3693

\bibitem[\protect\citeauthoryear{Yung, Somerville, Finkelstein, Popping  \&
  Dav\'{e}}{Yung et~al.}{2018}]{Yung_2018}
Yung L. Y.~A.,  Somerville R.~S.,  Finkelstein S.~L.,  Popping G.,   Dav\'{e}
  R.,  2018, \mn@doi [Monthly Notices of the Royal Astronomical Society]
  {10.1093/mnras/sty3241}, 483, 2983–3006

\bibitem[\protect\citeauthoryear{{Yung} et~al.,}{{Yung}
  et~al.}{2022}]{Yung22_semianalytic_lightcone}
{Yung} L.~Y.~A.,  et~al., 2022, \mn@doi [\mnras] {10.1093/mnras/stac2139},
  \href {https://ui.adsabs.harvard.edu/abs/2022MNRAS.515.5416Y} {515, 5416}

\bibitem[\protect\citeauthoryear{{Yung}, {Somerville}, {Finkelstein}, {Wilkins}
   \& {Gardner}}{{Yung} et~al.}{2023}]{Yung:2023bng}
{Yung} L.~Y.~A.,  {Somerville} R.~S.,  {Finkelstein} S.~L.,  {Wilkins} S.~M.,
  {Gardner} J.~P.,  2023, \mn@doi [arXiv e-prints] {10.48550/arXiv.2304.04348},
  \href {https://ui.adsabs.harvard.edu/abs/2023arXiv230404348Y} {p.
  arXiv:2304.04348}

\bibitem[\protect\citeauthoryear{{Zavala} et~al.,}{{Zavala}
  et~al.}{2023}]{Zavala_2023_interlopers}
{Zavala} J.~A.,  et~al., 2023, \mn@doi [\apjl] {10.3847/2041-8213/acacfe},
  \href {https://ui.adsabs.harvard.edu/abs/2023ApJ...943L...9Z} {943, L9}

\bibitem[\protect\citeauthoryear{{Zentner}, {Hearin}, {van den Bosch}, {Lange}
  \& {Villarreal}}{{Zentner} et~al.}{2019}]{Zentner_assemblybias}
{Zentner} A.~R.,  {Hearin} A.,  {van den Bosch} F.~C.,  {Lange} J.~U.,
  {Villarreal} A.~S.,  2019, \mn@doi [\mnras] {10.1093/mnras/stz470}, \href
  {https://ui.adsabs.harvard.edu/abs/2019MNRAS.485.1196Z} {485, 1196}

\bibitem[\protect\citeauthoryear{Zheng \& Weinberg}{Zheng \&
  Weinberg}{2007}]{Zheng:2005ef}
Zheng Z.,  Weinberg D.~H.,  2007, \mn@doi [Astrophys. J.] {10.1086/512151},
  659, 1

\bibitem[\protect\citeauthoryear{{van der Walt}, {Colbert}  \&
  {Varoquaux}}{{van der Walt} et~al.}{2011}]{numpy}
{van der Walt} S.,  {Colbert} S.~C.,   {Varoquaux} G.,  2011, \mn@doi
  [Computing in Science and Engineering] {10.1109/MCSE.2011.37}, \href
  {https://ui.adsabs.harvard.edu/abs/2011CSE....13b..22V} {13, 22}

\makeatother
\end{thebibliography}

\appendix

\section{Accretion Rates}
\label{app:accretionrates}

\begin{figure}
    \centering
\includegraphics[width=\linewidth]{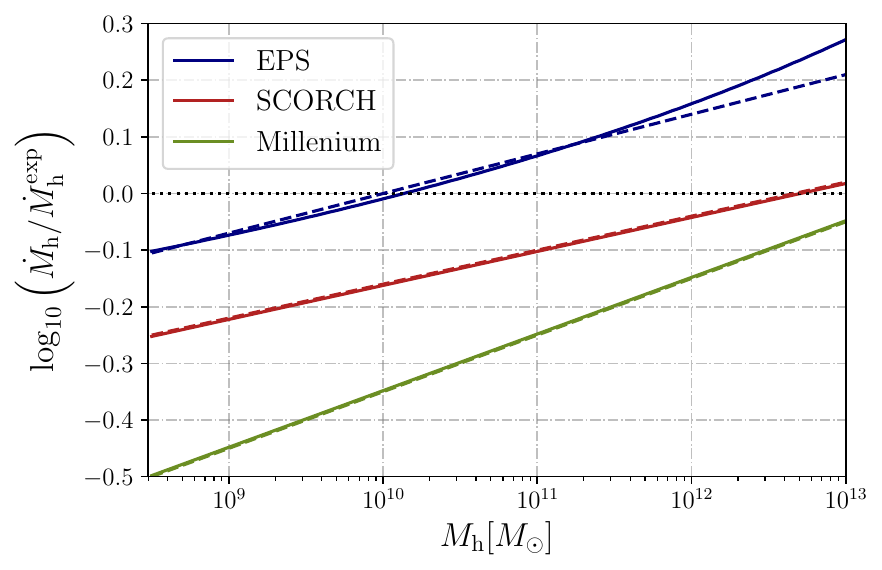}
    \caption{Accretion rates at $z=10$ calibrated to different simulations, normalized by the exponential model assumed in the main text. Dashed curves show (log-)linear fits, which match very well, meaning that different $\dot M_\mathrm{h}$ fits just translate into a shift of the $f_\star$ parameters (and provide equal predictions).
    }
    \label{fig:Mgdots}
\end{figure}

Our model assumes that the star-formation rate is the product of the (gas) accretion rate $\dot M_\mathrm{g} = f_\mathrm{b} \dot M_\mathrm{h}$ and the star-formation efficiency $f_\star$.
While there is no universal formula for the mass-accretion rate $\dot M_\mathrm{h}$ in the literature, here we argue that different functional forms are equivalent to each other, as their differences can be absorbed into the four parameters that determine $f_\star(M_\mathrm{h})$ (which are varied in our MCMCs).

Throughout the text, we assumed exponential accretion, where $\mathrm{d}M_\mathrm{h}/\mathrm{d}z = a_{\rm acc} M_\mathrm{h}$ and $a_{\rm acc}=0.79$, is calibrated to simulations~\citep{Schneider:2020xmf}. 
Other common prescriptions in the literature include the extended Press-Schechter formalism~\citep{Neistein:2006ak}, a fit to the {\tt Millenium} simulations~\citep{Fakhouri:2010st}, as well as a fit to the {\tt SCORCH} simulations~\citep{Trac:2015rva}.
We show these different accretion rates, divided by the exponential prescription, in Fig.~\ref{fig:Mgdots}.
This figure is at $z=10$, but is virtually identical at other $z$ in the range of interest.
For each curve, we present a linear approximation, which agrees remarkably well over the range of masses we consider.
This means that the differences with respect to the exponential model can be reabsorbed into the mass dependence of $f_\star$ (in particular the intercept of the linear fit will renormalize $\epsilon_{\star,\rm UV}$, and the slope will shift $\alpha_\star$ and $\beta_\star$ simultaneously).
As such, these different mass-accretion histories can be recast as a slightly different average SFE, otherwise providing identically good fits to all high-$z$ data.

\section{Stochasticity through a duty cycle}
\label{app:stochfduty}

\begin{figure}
    \centering
\includegraphics[width=\linewidth]{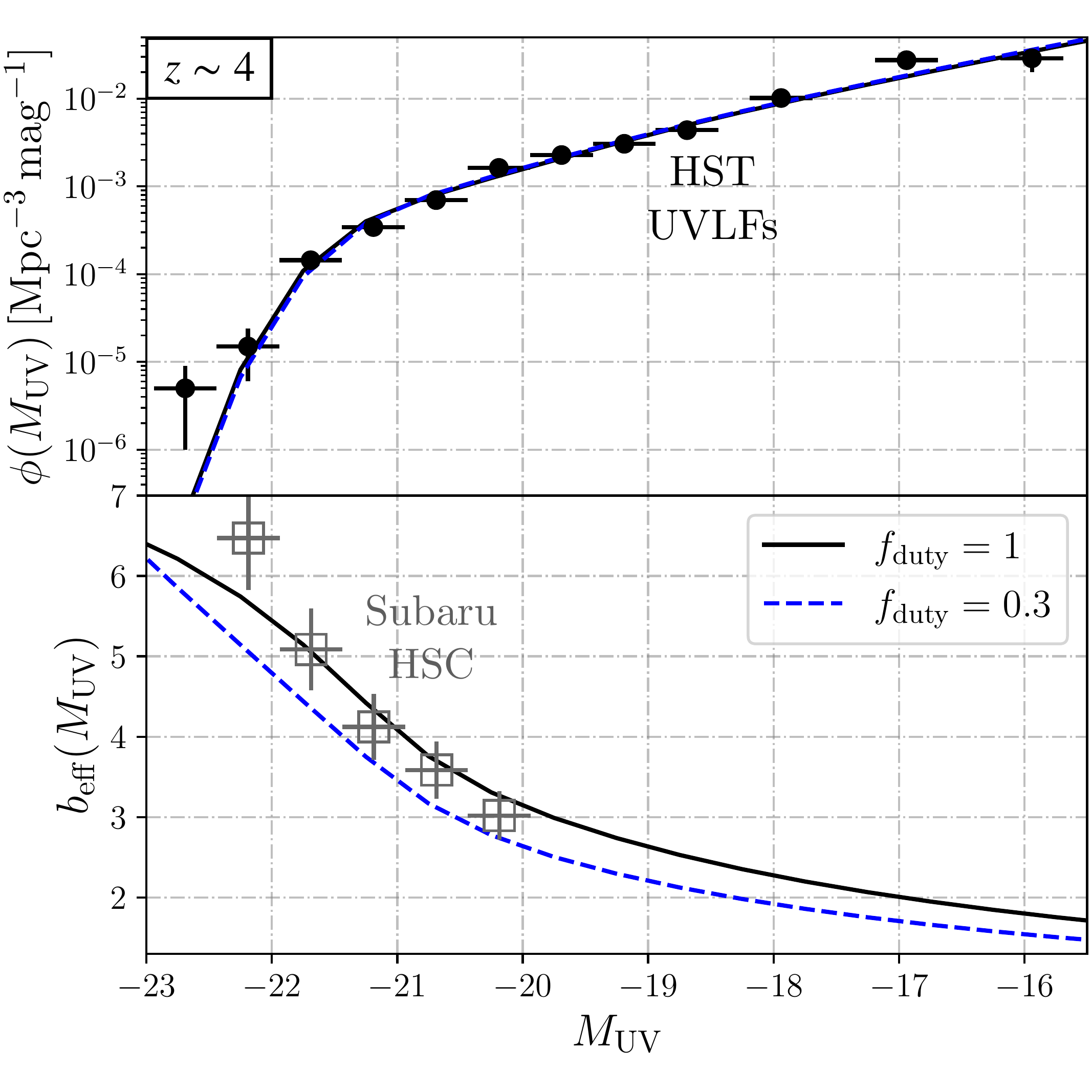}
    \caption{Same as Fig.~\ref{fig:z4degeneracy}, but for a model with a variable duty cycle $f_{\rm duty}$ (with fixed $\sigma_{\rm UV} = 0.3$). 
    Lowering $f_{\rm duty}$ can be compensated by adjusting the SFE $f_\star$ to provide a good fit to the UVLF, which however results in different bias predictions.
    }
    \label{fig:fduty}
\end{figure}

In the main text, we have focused on how stochasticity can manifest through the variance of the halo-galaxy connection $P(\MUV | M_\mathrm{h})$, given by:
\be
\label{eq:PMUVMh}
P(\MUV | M_\mathrm{h}) = \dfrac{1}{\sqrt{2\pi\sUV^2}} \exp\left[ - (\MUV - \overline{\MUV})^2/(2\sUV^2) \right]\ ,
\ee
with width $\sUV$ and mean $\overline{\MUV}(M_\mathrm{h})$.
A larger $\sUV$ ``blurs" the halo-galaxy connection, allowing low-mass halos to host luminous objects and vice-versa\footnote{
We note, in passing, that the average luminosity of galaxies shifts with $\sUV$, as a Gaussian distribution in $\MUV$ is log-normal in $L_{\rm UV}$, and thus will have a larger mean~\citep{Xavier:2016elr}. 
We have tested that the degeneracy remains after canceling this enhancement.}.

We also consider an alternative parameterization, where only a fraction $f_{\rm duty}\leq 1$ of galaxies are UV-bright at any point.
This can be achieved by reducing $P(\MUV | M_\mathrm{h})$ by a factor of $f_{\rm duty}$, and physically will occur whenever the duty cycle of galaxies is not unity.
We show in Fig.~\ref{fig:fduty} the UVLFs and bias predicted for two models, one with $f_{\rm duty}=1$ (as in the main text) and one with  $f_{\rm duty}=0.3$.
In the latter case, we have adjusted the rest of parameters in the halo-galaxy connection by hand to recover agreement in the UVLFs.
Despite the one-point functions being nearly identical, the biases predicted by these two models are significantly different.
The model with $f_{\rm duty}=0.3$ boasts a smaller bias, as each galaxy ought to be more luminous to account for the same UVLF (and thus will tend to reside in smaller mass halos).
{\it Subaru} HSC measurements mildly prefer $f_{\rm duty}\approx 1$ at $z=4$, though as in the main text, we warn the reader that the $b_{\rm eff}$ data from HSC cannot be directly compared against our predictions, as it was derived from an HOD model with a different cosmology.
This shows that clustering measurements are promising not only to determine $\sUV$, but also the duty cycle of galaxies.

\renewcommand{\thefigure}{D1}
\begin{figure*}
    \centering
\includegraphics[width=0.99\linewidth]{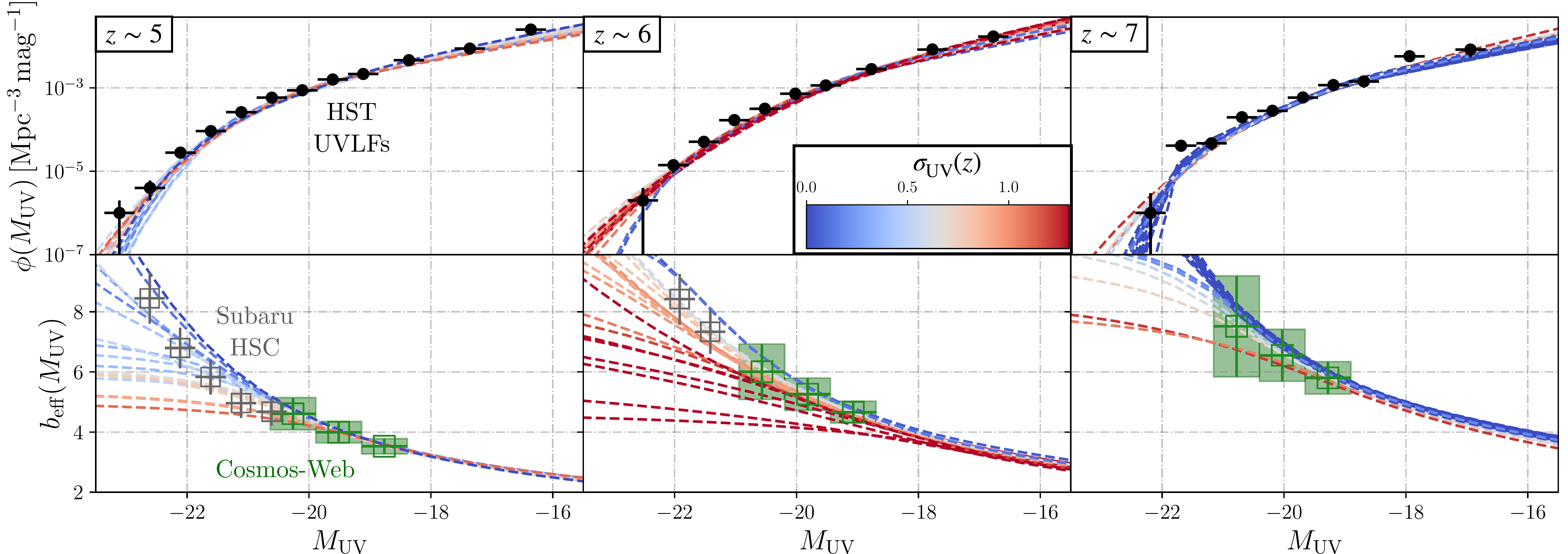}
    \caption{Same as Fig.~\ref{fig:z4degeneracy}, but for $z=5-7$. We have fit the astrophysical model at each $z$ independently, and show lines with parameters sampled from the 2$\sigma$ preferred region at each $z$, colored by their value of $\sUV$.
    Along with the reported biases from {\it Subaru} HSC (gray squares), we show the forecast for a Cosmos-Web-like JWST survey in green, which reaches deeper magnitudes and higher $z$.
    }
    \label{fig:UVLF_bias_z5to7}
\end{figure*}

\section{Bias data from HSC}
\label{app:biasHSC}

Here we describe how we derived the binned bias measurements from HSC that we show in this work.
These have been derived from the HOD analysis in~\citet{Harikane_Goldrush_2021}, which obtained $b_{\rm eff}(m_{\rm UV}<m_{\rm cut})$ for different UV magnitude cuts $m_{\rm cut}$. In order to translate these into binned (and number-weighted) biases, we take:
\be
b_{\rm eff}(m_{\rm UV}\in [m_{\rm cut,1},m_{\rm cut,2}]) = f_1 b_{\rm eff}(m_{\rm cut,1}) - f_2 b_{\rm eff}(m_{\rm cut,2})\ ,
\ee
for $f_i = n_i/(n_1-n_2)$, where $n_i$ are the number of objects in each cut-off from~\citet{Harikane_Goldrush_2021}.
Given the lack of error bars in $n_i$, and the fact that these biases are derived with an HOD model within a different fiducial cosmology, we always show the biases with a 10\% minimum error-bar.
In future work, we will use the full angular correlation function information from the {\it Subaru} HSC~\citep{Sabti_HSC_future}.

\section{Clustering at intermediate redshifts}
\label{app:z5to7}

In this appendix, we extend the analysis from Fig.~\ref{fig:z4degeneracy} to the $z=5-7$ range, each fit independently to the data from~\citet{Bouwens_2021_UVLFs}, see Fig.~\ref{fig:UVLF_bias_z5to7}.
Each $z$ suffers from the same SFE-$\sUV$ degeneracy that we studied in the main text, so their 2$\sigma$ confidence intervals cover a broad swath of values of $\sUV$. 
As we saw for the $z=4$ case, more stochasticity (larger $\sUV$) results in lower biases that flatten towards the bright end.
This is in conflict with the bias measurements from HSC~\citep{Harikane_Goldrush_2021}, which seem to disfavor $\sUV\gtrsim 1$ at $z\sim5$ and 6.
Moreover, we illustrate how a Cosmos-Web-like JWST survey can push the HSC results to fainter magnitudes and higher $z$.
This will be key to unveil the nature of the halo-galaxy connection at high redshifts.

\section{Best-fit parameters}
\label{app:bestfits}

In this appendix, we provide the best-fit values used throughout the text for the astrophysical parameters and outline our MCMC approach for ease of reproducibility.

We have used the {\tt emcee} code~\citep{Foreman-Mackey:2012any}, and run chains with the likelihood and priors described in the main text, where the UVLFs are predicted with the public {\tt Zeus21} code~\citep{Munoz:2023kkg}.
For each result, we have run $3.6\times10^5$ points (separated into 36 walkers), of which the first fifth are tossed out as burn-in.
Each point takes $\sim 1$ ms to run, so it takes a chain a few hours to converge.
We show our best-fit parameters for the four cases considered in the text in Tab.~\ref{tab:bestfits}.


\begin{table}
\begin{tabular}{lcccc}
                               & HST ($\epsilon_{\star,\rm UV}$) & +JWST   & HST ($\sigma_{\rm UV}$) & +JWST   \\ \hline\hline
$\alpha_\star$           & $0.61$                                  & 0.84    & 0.74                            & 0.69    \\
$\mathrm{d}\alpha_\star/\mathrm{d}z$             & $-0.01$                                 & 0.05    & 0.03                            & 0.01    \\
$\beta_\star$            & $-1.91$                                 & $-1.26$ & $-1.76$                         & $-1.68$ \\
$\mathrm{d}\beta_\star/\mathrm{d}z$              & $0.08$                                  & 0.14    & $-0.02$                         & 0.18    \\
$\log_{10}\!M_c$           & $12.03$                                 & 11.83   & 11.84                           & 11.93   \\
$d\log_{10}\!M_c/dz$             & $0.03$                                  & $-0.03$ & $-0.02$                         & $0.00$  \\ \hline
$\log_{10}\epsilon$ & $-$                                     & $-$     & $-1.08$                         & $-1.11$ \\
$\mathrm{d}\log_{10}\epsilon/\mathrm{d}z$  & $-$                                     & $-$     & $-0.07$                         & $-0.08$ \\
$\sigma_{\rm UV}$         & 0.65                                    & 0.59    & $-$                             & $-$     \\
$\mathrm{d}\sigma_{\rm UV}/\mathrm{d}z$          & $-0.03$                                 & $-0.03$ & $-$                             & $-$    \\
\hline
\end{tabular}
\caption{Best-fit values for the astrophysical parameters used in the main text, as defined in Eq.~\eqref{eq:fstar}. Results are shown for two models, one in which we vary $\epsilon \equiv \epsilon_{\star,\rm UV}$ at each $z$ (left two columns, reported in blue in Fig.~\ref{fig:posterior}) and one in which we do the same for $\sUV$ (right two columns, red in Fig.~\ref{fig:posterior}); for an analysis with HST only or HST + JWST data.
The rest of parameters are assumed to vary linearly with $z$ in each case, and we report their values and derivatives at $z=8$.
}
\label{tab:bestfits}
\end{table}

\end{document}